\documentclass[10pt]{article}
\pdfoutput=1 

\usepackage[top=2.0cm, bottom=1.5cm, inner=2.0cm, outer=2.0cm]{geometry}
\usepackage{float}
\usepackage{enumerate}
\usepackage{color,graphicx,colortbl}
\usepackage{paralist,parskip}
\usepackage{booktabs, multirow, makecell}
\usepackage{subfigure, bm, comment}
\usepackage{amsmath,amsfonts,amssymb}
\usepackage{mathabx,fixmath,algorithm, amsthm} 
\usepackage{pstool}
\usepackage{psfrag}
\usepackage{hyperref}
\usepackage{siunitx}
\usepackage{cite}
\usepackage{algpseudocode} 
\usepackage{bm} 
\usepackage[table,xcdraw]{xcolor} 

\def\lecs{L$\epsilon$-CS} 
\definecolor{Gray}{gray}{0.9}
\newcommand{\mul}[2]{$\begin{array}{c} #1 \\ #2 \end{array}$}

\newtheoremstyle{spaced}
  {8pt}
  {2pt}
  {\itshape}
  {}
  {\bfseries}
  {.}
  {2pt}
  {}
  
\theoremstyle{spaced}

\newtheorem{theorem}{Theorem}

\newcommand{\rev}[1]{#1}

\newcommand{\revision}[1]{#1}

\newcommand{\Centr}{$\texttt{Centralized}$}
\newcommand{\Coop}{$\texttt{Cooperative}$}
\newcommand{\Comp}{$\texttt{Competitive}$}

\usepackage{natbib}
\usepackage{csquotes}
\usepackage{cleveref}

\title{On the needs for MaaS platforms to handle competition in ridesharing mobility}

\author{Venktesh Pandey, Julien Monteil, Claudio Gambella, and Andrea Simonetto
\thanks{Venktesh Pandey is with the University of Texas at Austin, {\tt\small \{$venktesh@utexas.edu$\}}. Julien Monteil, Claudio Gambella, and Andrea Simonetto are with IBM Research, Ireland Lab {\tt\small \{$julien.monteil@ie.ibm.com$\}}, {\tt\small \{$claudio.gambella1@ie.ibm.com$\}}, {\tt\small \{$andrea.simonetto@ibm.com$\}}. All authors contributed equally to the paper.}
}
\date{}

\begin{document}

\pagenumbering{arabic} 
\pagestyle{plain}

\maketitle

\begin{abstract}

Ridesharing has been emerging as a new type of mobility. However, the early promises of ridesharing for alleviating congestion in cities may be undermined by a number of challenges, including the growing number of proposed services and the subsequent increasing number of vehicles, as a natural consequence of competition. In this work, we present optimization-based approaches to model cooperation and competition between multiple ridesharing companies, in a real-time on-demand setting. A recent trend relies on solving the integrated combination of Dial-A-Ride Problems (DARP), which compute the cost of assigning incoming requests to vehicle routes, plus Linear Assignment Problems (LAP), which assign vehicles to requests. While the DARPs, are solved at the level of the vehicles of each company, we introduce cooperative and competitive approaches to solve the LAP. The cooperative model, which could make use of Mobility as a Service platforms, is shown to solve the LAP to optimality following closely results from the literature, and limiting the amount of information the companies are required to share. We investigate how a realistic model of competition deviates from this optimality and provide worst case bounds. 
We evaluate these models with respect to a centralized model 
on one-week instances of the New York City taxi dataset. Model variants coping with noise in the travel time estimations, bias in the assignment costs, and preferences in the competitive case are also presented and validated. 
The computational results suggest that cooperation among ridesharing companies can be conducted in such a way to limit the degradation of the level of service with respect to a centralized model. Finally, we argue that the competition can lower the quality of the ridesharing service, especially in the case customer preferences are accommodated.
\end{abstract}



	

\section{Introduction}\label{sec:intro}

Dynamic ridesharing and \revision{ridesourcing} have gained significant traction in recent years~\citep{agatz2012optimization, MOURAD2019} as on-demand mobility services for transporting customers from one city location to another. For US cities only, it is estimated that 2.61 billion passengers were transported by these services in 2017, reporting a 37$\%$ increase compared to 2016~\citep{schaller}.
\revision{Ridesourcing provides prearranged and on-demand transportation services where drivers of personal vehicles are matched with the passengers and receive compensation \citep{jin2018ridesourcing}. Ridesourcing services are also referred and associated with several other names including transportation network companies \citep{PRATT2019459}, real-time ridesharing \citep{MASOUD2017218, VIVODA2018426}, ride-hailing \citep{XU2019, ALEMI2019233, YOUNG2019383, CONTRERAS201863}, e-haling \citep{HE201593} and on-demand rides~\citep{rayle2016just}.
	 The founding concept of ridesourcing is that a customer hires a driver to take her/him to a specific destination, characterizing itself as a personal transportation experience \citep{LAVIERI2019100}. Known possible drawbacks of ridesourcing are the problem of empty vehicles-miles-traveled and congestion phenomena \citep{NIE2017242}.	
	As a possible attempt to mitigate this, ridesharing, also referred as carpooling or vanpooling, facilitates shared rides between drivers and passengers with similar origin-destination pairings, and allows travelers going from one destination to the other to pick up other passengers on the way~\citep{shaheen2016shared, COULOMBEL2019110}. 	
	Scheduling routes that are matching customers' preferences and needs requires to consider a specific class of vehicle routing problems, namely the dial-a-ride problems \citep{ho2018survey, FU2002291}. In dial-a-ride services, the provider aims to minimize operations costs or level of service, while satisfying demand and customers' constraints \citep{Cordeau2003, parragh2008survey}.	 Addressing the human end economic perspectives in large-scale network increases the problem complexity, and necessitates tailored solution approaches \citep{MUELAS2015110, BONGIOVANNI2019436} and taking conflicting objectives into account \citep{GUERRIERO2014299, markovic2015optimizing}.
	Dial-a-ride problem arise in transport services such as paratransit, or demand responsive transit, for mobility-impaired individuals \citep{NGUYENHOANG2010841, DEKA2014181, PHUN2018175}. Paratransit enhances the public transport means with individualized trips and door-to-door services; it involves challenging implementation problems, specifically the scheduling of routes in large-scale networks \citep{DIKAS201418, KARABUK2009448} , the integration in the existing traffic levels \citep{PHUN2018175}, evaluation of operational costs \citep{SCHALEKAMP201758, FITZGERALD2000261, GUPTA2010201} and coping with uncertain demand \citep{BEARSE2004809} and traffic conditions  \citep{FU2002485}.

	 In this research, we focus on the ridesourcing services, including the possibility of ridesplitting \citep{LI2019330}, meaning to encompass on-demand shared-mobility services. We adopt the term \textit{ridesharing} to refer to this kind of services: this helps to position this paper in line both with established Transportation literature for on-demand shared-mobility services (see, e.g, \cite{alonso2017demand, simo2018demand, CHEN201751, WANG2019390, LEI2019, DI2018230}), and the recent tendency of ridesourcing providers (e.g., Uber and Lyft) to offer ridesplitting/carpooling services and therefore entering in the ridesharing space \citep{SCHWIETERMAN20189, MOODY2019258}. The possibility that passengers have to share vehicle routes is heavily exploited in the design of the services proposed in this paper, by dynamically inserting pickup and dropoff requests in the vehicle schedules.}



The effects of ridesharing services are mixed: on the one hand, they offer a convenient and reliable way to meet travel demand, and enable to increase car occupancy \citep{fiedler2018impact}, but on the other hand they can also lead to increased congestion when there is excess supply of vehicles. It is estimated that ridesharing mobility added 5.7 billion miles of driving~\citep{schaller} in the metropolitan areas of Boston, Chicago, Los Angeles, Miami, New York, Phildelphia, San Francisco, Seattle, and Washington DC alone.
An ever-growing number of private ridesharing companies (e.g., Uber, Lyft, Didi Chuxing, and Via) participate in the ridesharing market. These companies naturally compete for the same set of resources (i.e., customers) to gain a significant market share. In an unregulated environment, this competition certainly leads to non-optimal behaviors, in terms of the number of vehicles present on the roads, and of the quality of the service achieved. As indicated by the NYC Taxi and Limousine Commission Summary Reports~\citep{NYCreports}, the number of daily trips for Taxis, Uber, and Lyft has changed from roughly 500k, 80k, and 0k in the first quarter of 2015 to 300k, 500k, and 140k in the last quarter of 2018. At the same time, the same reports find that while the number of taxis has decreased from about 20k to 15k, the number of vehicles for Uber has increased drastically from 12k to 76k, and for Lyft went from 0k to 46k (see~\cite{todd-report18}). Therefore, it is clear that the two main ridesharing companies, Uber and Lyft, have supplanted the New York taxi company, and that their growth is still ongoing. This raises questions about the fairness of the competition and the level of congestion that such growth may be causing.

The development of large-scale ridesharing solutions, dealing with thousands of requests and vehicles in real-time, is an emerging and challenging topic of research. Very recent research investigates the advantages of introducing meeting points in terms of cost savings and congestion mitigation~\citep{StiASG15}, the consideration of riders' satisfaction and privacy rights~\citep{AivGHK16}, the integration of ridesharing in multi-modal systems~\citep{YAN2018, LIU2018}, the offering of tailored pricing schemes \citep{SAYARSHAD2018192} (e.g., for regular  such as commuters~\citep{LiuL17, MaZ17}), the study of the changes in travel patterns induced by such ridesharing systems~\citep{DonWLZ18},  the evaluation of performance metrics~\citep{NAJMI2017122},  and the consideration of ridesplitting as a binary classification problem~\citep{chen2017understanding}.

In this paper, we consider as a building block the works of~\cite{alonso2017demand,simo2018demand}, which provide solutions for solving the \textit{centralized} real-time city-scale ridesharing problem, by mapping incoming batches of requests with available vehicles, in a three-step procedure: (i) selecting candidate vehicles to serve requests, based on vehicle location and seat occupancy; (ii) computing assignment costs incurred for serving a request with a selected vehicle, while meeting customer constraints, by solving a Dial-a-Ride problem (DARP)~\citep{Cordeau2006,HAME201111,LIU2015267};(iii) performing optimal assignments of requests to vehicles, given the assignment costs determined in step (ii). In~\cite{simo2018demand}, it is highlighted that linear assignments can perform as good as more elaborated assignments, when run at a high enough sampling rate. \cite{simo2018demand} shows that optimization-based approaches for on-demand ridesharing can provide a high level of service while limiting the growth of the fleet size. Specifically, it is shown that $98\%$ of the current trip demand in Manhattan, available via the New York Taxi dataset~\citep{NYCdata}, can be satisfied with only 20$\%$ of the current taxi fleet~\citep{alonso2017demand}. In a similar fashion, the work of~\cite{LOKHANDWALA201845} argues that the taxi fleet size can be reduced by 59$\%$ if shared autonomous taxis are used while considering individual customer preferences. Such large-scale ridesharing solutions propose a centralized way of dealing with customer requests, which requires a central coordinating agent or \textit{broker}. However, in practice, there is no such centrality, as every customer will book a ride via the proprietary smartphone app of a given company. In fact, companies will compete for accommodating as many customers as possible, and the overall behavior of the system may be far from the centralized one.

The consideration of decentralization due to the multiple actors of the transportation network, which may involve some degree of cooperation and competition between these actors, is often overlooked, as noted by~\cite{wang2017game}. In fact, as for ridesharing mobility, few works focus on non-centralized systems. A first work focusing on the taxi industry is the one of~\cite{cairns1996competition} which shows that regulation of fares and fleet sizes is essential in a multi-company competition context. In another pioneering work~\citep{yang2002demand}, a demand-supply equilibrium of competitive taxi services is formulated based on a network model. The model can then be used for the regulation of taxi fleet sizes and fares across companies. In the recent work of~\cite{qian2017taxi}, the taxi market is modelled as a multiple leader-follower game, where the leaders are the passengers, {while the traditional and third party taxi services are the followers. An approximate Nash equilibrium is proved to exist for the model and the fleet size and pricing policy are shown to be tied to the level of competition, which in turn has several impacts in terms of the quality of the service achieved.
These decentralized models rely on the assumption that companies are willing to share information on their vehicles (e.g., location). This may not be a realistic requirement for competing ridesharing providers.

In this work, we aim at providing a first analysis to quantify the effects of competition between ridesharing companies at a city-scale. This effort fits well into the ongoing reconsideration of mobility services, including the emerging Mobility as a Service (MaaS) business models~\citep{goodall2017rise}. The recent contribution of \citep{Sejourne2018} quantifies the rebalancing burden caused by the demand fragmentation between multiple companies, but neglects competition forms. In the context of the above literature, we move away from the centralized city-scale ridesharing solution and propose a cooperative and a competitive approach, which consider the current  multi-company ridesharing landscape. The proposed \texttt{Cooperative} model does not require companies to share any proprietary information about their fleet, and may be well suited to the use of Mass platforms. The \texttt{Competitive} model does not require any information sharing in-between companies: it first corresponds to the situation of customers booking rides via a broker that provides the best offer among the companies, and it also enables to evaluate the broker-free situation where customers book rides via one ridesharing company app on their smartphones.

%


In order to evaluate advantages and limitation of the \Coop~and \Comp~models, we compare them with a \Centr~model, obtained by adapting the work of~\cite{simo2018demand} to the multi-company setting. Hence, a suitable proxy for an efficient \texttt{Centralized} model for multi-company ridesharing systems is as follows: (i) the system collects batches of customer requests for a given time period, e.g., 10~\si{\second}, (ii) the companies compute the costs for inserting requests in each vehicle route, by solving a DARP problem, (iii) those costs are shared with the broker, (iv) the broker solves the Linear Assignment Problem (LAP) to optimality and sends the updated assignments of vehicles to requests (v) the broker adopts a rebalancing strategy for the unserved requests, which means running steps (i)-(iv) with loose time constraints. This process is repeated for every batch of requests. In particular, in~\cite{simo2018demand}, the authors highlight that one-to-one assignments of customer requests to vehicles provide very high service rates in a dynamic context, and suggest that a myopic search for optimal multiple-to-one assignments of customer requests to vehicles is not necessary\revision{, and not beneficial from the computational point of view}. In this context, the centralization of the ridesharing solution comes from the need for a broker that collects the costs and solves a centralized LAP.

In this work, we put emphasis on ridesharing services where the companies participating in the city-scale ridesharing demand and supply process either share limited information or do not share information with a central broker. In these settings, the implementation of a centralized communication protocol is not possible. To our knowledge, this is the first work to bring insights on the quality of the ridesharing services in such realistic communication and information sharing settings.


The contributions of this work are the following:
\begin{enumerate}[{\textbullet}]
	\item We propose a \texttt{Cooperative} model of ridesharing that relies on a broker receiving only limited information from all companies. Variants of this model include the consideration of noise, i.e. the bids shared with the broker are not exact, and bias, i.e. a company may systematically under-, respectively over-estimate the costs it shares -- the under-estimation may be even wanted as a proxy to discount costs and appeal to a larger number of customers. 
	\item We prove that this \texttt{Cooperative} model, in the absence of noise and bias, achieves optimal assignment solutions, i.e. its solution is the same as the centralized approach. Our result is based on the work of~\cite{Naparstek2014}.
	\item We propose a \texttt{Competitive} model that aims at capturing the current status of ridesharing in cities, where companies route their vehicles to serve as many requests as possible, and each customer may choose the best offer across these non-cooperative companies. A very relevant variant of the model is the consideration of customer preferences, where the customers may prefer to be served by a certain company, even if another company would offer a lower cost. Preferences can be flexible and ignored, if the difference between the cost offered by best company and by the favorite one is lower than a certain threshold, or they can be strict if, e.g., customers book rides through the proprietary app of a given company and only have access to the offering of that company. We consider both these cases and a mix thereof. 
	\item We prove that the \texttt{Competitive} model degrades the centralized optimal cost by a factor 2 in the worst case when two companies are involved, and by a factor 3 in the worst case when more than two companies are involved. 
	\item We test the performance of the cooperative and competitive models and their variants on different cost matrices, in a static setting, which confirms the previous results and allow us to draw first conclusions regarding the influence of the protocols on the gap to optimality.  
	\item We conduct tests on large-scale ridesharing instances, using the NYC Taxi dataset~\citep{NYCdata}. 
	and we provide recommendations on ridesharing control policies, in particular in the context of MaaS platforms. 
\end{enumerate}
		
The remainder of this article is organized as follows. Section~\ref{sec:federated} presents the overall architectures for \texttt{Centralized}, \texttt{Cooperative}, and \texttt{Competitive} models. 
Section~\ref{sec:coop} and Section~\ref{sec:comp} present a theoretical analysis on the \texttt{Cooperative} and \texttt{Competitive} models, respectively, with results proving optimality for the cooperative case, and bounds on the gap to optimality for the competitive case. Section~\ref{sec:num} reports the computational results on the New York City taxi fleet data set, for the different considered protocols and their variants. We conclude in Section~\ref{sec:conclusion} with a summary of our findings and future research perspectives.



\section{Models and architectures for dynamic ridesharing algorithms}	\label{sec:federated}

We present here the models and architectures for the dynamic ridesharing algorithms. We start from the centralized algorithm put forward in~\cite{simo2018demand}, which serves as a stepping stone to investigate cooperative and competitive approaches for multi-company ridesharing. 
One key feature of~\cite{simo2018demand} is that the optimal assignment problem is reduced to a Linear Assignment Problem (LAP), which scales very well in terms of number of requests and fleet size. This reduction is possible by the design assumption that only one new request can be assigned to a  given vehicle per sampling period. We call such assignments single-request assignments. As shown in~\cite{simo2018demand}, on one hand, if the sampling period is small enough, then single-request assignments are not detrimental to service quality; on the other hand, this assumption unlocks important computational and algorithmic advantages, e.g., lower computational effort, and the possibility to run the algorithm in a distributed fashion, which will be further exploited here.  

\subsection{Preliminary: Ridesharing \rev{logic}}\label{sec:prel}

\revision{
We formalize here the ridesharing problem under consideration. The centralized ridesharing logic is the one considered in \cite{simo2018demand} for the single-company case. The reader is referred to \cite{simo2018demand} for detailed discussions. 

Let $\mathcal{M}$ denote a set of customer trip requests at time $t$ and $\mathcal{P}$ denote a set of companies. For a given batch of requests, each company $p \in \mathcal{P}$ has a fleet of available vehicles, denoted by set $\mathcal{C}_p$, which are available for the customers to request. The set $\mathcal{C}$ contains all the available vehicles, i.e., $\mathcal{C} = \bigcup_{p} \mathcal{C}_p$. Let $p_i$ denote the company to which a vehicle $i \in \mathcal{C}$ belongs. Each vehicles has its own seat capacity. The ridesharing problem consists in providing a real-time assignment of requests (customers) to available vehicles and their correspondent routes, while minimizing the total travel times of vehicles. Available vehicles are those that can pick up customers, while complying with the time constraints associated with the requests, and without exceeding their seat capacity.

The ridesharing logic is run at specific time instants $t_k$ ($k =0,1,2,\dots$):
\begin{enumerate}
	\item It obtains the customer requests submitted during the time interval $[t_{k-1}, t_k)$.
	\item It sends requests to context mapping module, which filters the closest vehicles to each customers pick-up point (see Section 3.2. of \cite{simo2018demand}). 
	\item It asks customer assignment costs to the vehicles filtered by the context mapping module, by solving a Dial-a-Ride Problem (DARP). 
	\item It calls an optimization module to determine optimal assignment of requests to vehicles. \cite{simo2018demand} showed that, if the sampling window $[t_{k-1}, t_k)$ is small enough, then the assignment of requests in batches leads to minimal losses of service quality (see Section 4.2).
	\item It sends the assignments and their corresponding routes to customers and vehicles. 
	\item If some customers cannot be served, it calls an internal rebalancing module, which runs the logic again from (2. to 5.) with relaxed time preferences and for idle vehicles (i.e., without passengers) only (see Section 3.5. of \cite{simo2018demand}). 
\end{enumerate} 

The ridesharing logic involves two optimization problems: a DARP which estimates the assignment costs of requests to vehicles, and an assignment problem to allocate the requests to available vehicles, at minimum travel time.

\paragraph{Dial-a-Ride Problem.} At the level of each company, and independently of the communication protocol in place, every time a new batch of customers is processed, each vehicle $i$ is required to  estimate the time duration of the route that serves the already scheduled customers and each new customer $j$. This is accomplished by solving a single-vehicle Dial-a-Ride-Problem (DARP) \cite{HAME201111, LIU2015267, Ropke2007}, with the aim of minimizing the route duration. In particular, the input for DARP for vehicle $i$ is: \emph{(i)} the customers to be served in the future, consisting of the new request $j$, and the scheduled customers in the current route $R_i$ (determined in the previous optimization run), \emph{(iii)} pickup and delivery locations and time limitations/preferences of the scheduled customers, \emph{(iv)} current location $M_i$ of the vehicle and its capacity $C_i$, and \emph{(v)} the matrix $\tau$ of the travel times between pick-up/delivery locations, location $M_i$ and request $j$. The output of the DARP is expressed as:
\begin{equation}
(R_{ij}, c_{ij}) = \mathrm{DARP}(j, R_i, M_i, C_i, \tau), \label{prob:DARP}
\end{equation}
where:
\begin{itemize}
	\item $R_{ij}$ is a route for vehicle $i$, with a pick-up and drop-off schedule that serves request $j$ and the customers scheduled in route $R_i$
	\item $c_{ij}$ is the time duration of $R_{ij}$.
\end{itemize}
 The DARP \eqref{prob:DARP} is solved via an insertion heuristic (formalized in Algorithm 1 by \cite{simo2018demand}), which evaluates the extra travel time that vehicle $i$ incurs to when adding request $j$ in a feasible position in the scheduled route $R_i$. In our implementation, the first $3$ feasible insertions of request $j$ in route $R_i$ are evaluated and the best of the $3$ (in terms of travel time) corresponds to the new route $R_{ij}$ in equation \eqref{prob:DARP}. Being a constructive heuristic, this solution approach does not require to consider subtour elimination constraints. We remark that, for each request $j$, only the vehicles filtered by the context mapping module are required to estimate the insertion cost: this limits the number of single-vehicle DARPs to solve in each optimization run. Under the assumption that only one request can be assigned to the previously scheduled routes, multiple new requests are not interfering in the DARP of a fixed vehicle in a given time instant.

   In the ridesharing optimization, $c_{ij} \in \mathbb{R}_+$ is the cost for assigning vehicle $i$ to request $j$ required in Step 4 of the logic. 
   
 \paragraph{Assignment Problem.} As described in Steps 4 and 6, a linear assignment problem, is solved twice in each sampling period: first, to seek an optimal assignment of requests to available vehicles, and then to rebalance vehicles to satisfy unserved requests, by relaxing time preferences specified in the requests.
}
Let $\mathcal{M}$ denote a set of customer trip requests at time $t$ and $\mathcal{P}$ denote a set of companies. For a given batch of requests, each company $p \in \mathcal{P}$ has a fleet of available vehicles, denoted by set $\mathcal{C}_p$, which are available for the customers to request. The set $\mathcal{C}$ contains all the available vehicles, i.e., $\mathcal{C} = \bigcup_{p} \mathcal{C}_p$. Let $p_i$ denote the company to which a vehicle $i \in \mathcal{C}$ belongs. Each vehicles has its own seat capacity and can accommodate multiple customers at any time. 
 
Having computed the assignment costs, let $x_{ij} \in \{ 0,1\}, i \in \mathcal{C}=\{1, \dots, n\}, j \in \mathcal{M}=\{1, \dots, m\}$ be the set of binary variables: $x_{ij}=1$ only if vehicle $i$ is assigned to customer $j$, otherwise $x_{ij}=0$. We further denote by $j_i$ the customer $j \in \mathcal{M}$ assigned to a vehicle $i \in \mathcal{C}$, that is, $x_{i,j_i}=1$.


For our optimization structure, in each sampling step we seek to assign each vehicle to exactly one new customer. In other words, vehicles can be assigned to multiple customers over successive time steps, but never within the time step. This is formalized by the following linear assignment problem: 
	\begin{subequations}
 		\begin{align}
	 		\min_{x_{ij} \in \{ 0,1 \}, i\in \mathcal{C}, j \in \mathcal{M} } & \sum_{i=1}^n \sum_{j=1}^m c_{ij} x_{ij} & \label{eq:optimization} \quad \left[  \textbf{LAP(}\mathcal{C},\mathcal{M} \textbf{)} \right]\\
	 		\text{subject to: } & \sum_{i=1}^n x_{ij} = 1, & j \in \mathcal{M} \label{eq:const1}\\
	 		 & \sum_{j=1}^m x_{ij} = 1, & i \in \mathcal{C} \label{eq:const2}
 		\end{align}
 	\end{subequations}
where, without loss of generality, we assume that the number of vehicles is equal to the number of new customers, $n=m$. Note that if this is not the case, one can augment the vehicle set or the customer set with dummy vehicles/customers with infinite assignment costs.

The goal of the optimization is to minimize the total assignment costs \eqref{eq:optimization}: in our implementation, $c_{ij}$ is expressed as the time duration (or TD) of the route of vehicle $i$ that serves the already scheduled customers and the new customer $j$. The solution of the assignment problem $\text{LAP}(\mathcal{C},\mathcal{M})$ needs then to be interpreted for practical feasibility, given the value of the assignment costs. If an infinite cost $c_{ij}$ is in solution, then vehicle $i$ will not insert request $j$ in the route, and customer $j$ would be a candidate for the relocation phase.  Provided that all the costs $c_{ij}$ are precomputed, then (\ref{eq:optimization}-\ref{eq:const2}) can be casted as a symmetric LAP \citep{KenH80, PapS82, bijsterbosch2010solving}. It is well known that LAPs have a totally unimodular constraint matrix, hence their continuous relaxation (i.e., substitute $x_{ij}\in\{0,1\}$ with $x_{ij}\in[0,1]$) is exact. We exploit the available efficient solution algorithms for LAPs, in particular the auction algorithm (scaling up to $10^6$ requests), see e.g.~\citep{Bernard2016, bertsekas1990auction}.


\subsection{The three models of multi-company ridesharing}\label{3-model}

In this section, we give further detail on the three models of interaction between ridesharing companies, ie., the \Centr, \Coop, \Comp~models, along with their variants of practical interest: 
\begin{enumerate}
\item[\textsf{A}-] The \Centr~model where a broker (or central authority) collects all the customer requests, asks the companies to compute the costs associated with assigning their vehicles to any customer, receives their costs, which are then collected in a linear assignment problem \textbf{LAP(}$\mathcal{C},\mathcal{M}$\textbf{)}, which it solves. The LAP can be solved using efficient and scalable algorithms, e.g., centralized auction algorithms~\citep{bertsekas1990auction}.

\item[\textsf{B}-] The \Coop~model where the broker does not require access to the matrix of all the costs $c_{ij}$ for every vehicle-request pair (from which it could reconstruct company-private information, e.g., vehicle locations in the \Centr~ model), instead the broker interacts with each company by iteratively asking for the best prices for the yet unassigned customers. This model, which is essentially an extension of the distributed auction algorithm of~\cite{Naparstek2014}, runs until all customers are assigned to vehicles. It relies on a distributed architecture, and it minimizes information sharing (no proprietary information of the companies is shared) but, as we shall prove, still enables to solve \textbf{LAP}$(\mathcal{C},\mathcal{M})$ to optimality. Extensions of this protocol introduce stochastic noise in the computation of the $c_{ij}$'s, which can come from the vehicle positioning errors and the routing computations, and bias as the routing/map service may be different among the companies and a given routing service may provide consistently shorter/longer travel times compared to others, or a company may use discounted costs to attract more customers. Models of cooperation have been investigated in applications such as logistic shipping (see \cite{IRANNEZHAD2018312}).

\item[\textsf{C}-] The \Comp~model where there is no broker to coordinate the assignment process, and customers send their requests either to a web-service (which finds the best deal) or to a company application on their smartphone. Each company $p$ solves a company-wide \textbf{LAP(}$C_p, \mathcal{M}$\textbf{)} and submit potential assignments directly to the web-service or to the customer apps. The customers who receive an offer (either through the web-service or to the app) then select the best offer among the companies (considering them rational agents) and are considered assigned. Each company then re-solves the assignment considering all unassigned customers. The process continues until all customers are assigned. An extension of this protocol is to consider customer's preferences. Specifically, each customer is willing to pay an extra cost $\gamma_j$ to his preferred ridesharing company $i^*$ even if another company would offer a lower cost. Preference can be flexible and ignored, if the difference between the cost offered by best company and by the favorite one is lower than a certain threshold, or they can be strict if, e.g., customers book through the proprietary app of a given company and only have access to the offer of this company, in which case each company $p$ solves a company-wide \textbf{LAP(}$C_p, \mathcal{M}_p$\textbf{)}. Noise and bias can be incorporated in the \Comp~model, as well.
\end{enumerate}

We argue that the \Centr~ protocol is not suitable for the practical implementation of multi-company ridesharing. This is because each company-owned vehicle would need to communicate directly to the broker the assignment costs. Hence, each company would disclose proprietary information with the broker. For this reason, the scope of paper is to analyze the solutions of the \Coop~and \Comp~models.


In Figure~\ref{fig:architectureA}, we present the \Centr~architecture. The requests are collected in batches of few seconds by a broker which lives on the public/city cloud or it is provided via a MaaS platform. The broker sends the requests to all the registered companies. 
Each company cloud, denoted $p\in\mathcal{P}$, consumes \revision{at time instant $t_k$ the entire group of requests arrived in interval $[t_{k-1}, t_k)$}. First it runs a context mapping algorithm, which consists in filtering the vehicles that could serve the new customers, given their mutual geographic location. In other words, there is no reason to ask a vehicle on a route that is on \emph{that} side of the city to see whether it could accommodate a new customer on \emph{this} side of the city. This module considerably reduces the communications \rev{and computational} requirements\revision{: in other words, if $J$ requests have arrived, and $N_p$ is the number of vehicles $i_p$ that can serve a request $j$ upon applying context mapping of company $p$, then  at most $J \cdot N_p$ DARPs are solved in each optimization run. As will be discussed in Section \ref{sec:num-real}, the number $N_p$ is set to $10$ for every company $p$. However, the centralized architecture could also be able to cope with company clouds with different context mapping capabilities.}. Second, \revision{each company} asks its selected proprietary vehicles to compute the cost $c_{ij}$ of incorporating customer $j$ into the schedule of vehicle $i$ (this is done solving a DARP with possible time-windows constraints, as described in Section \ref{sec:prel}). Third, the company cloud sends the collection of $c_{ij}$ for all the requests $\mathcal{M}$ and all its vehicles $\mathcal{C}_p$ to the public cloud. The broker collects all the costs $c_{ij}$ coming from all the companies and builds the \textbf{LAP(}$\mathcal{C},\mathcal{M}$\textbf{)}, which it is solved on the public cloud and determine the final vehicle-to-customer assignment and schedule. This information is sent both to the customers and to the companies. 

The unmet requests are then reprocessed in the same way. The broker sends to the company clouds the unmet requests flagged as unmet. The companies run the context mapping with a much looser search radius and ask their selected vehicles to determine the cost of accommodate the requests, with much looser constraints (e.g., no time windows). The costs are collected and then sent to the broker who instantiate another linear assignment problem. The assignments that result from the LAP may not be suitable for the customers, who can refuse the match, or accept it at a lower price. \revision{The reason to run the rebalancing is that idle vehicles will move towards areas where the demand is higher.} We refer the reader to~\cite{simo2018demand,alonso2017demand} for further details on this rebalancing strategy.

\begin{figure}[H]
	\centering
	\includegraphics[width=0.6\textwidth]{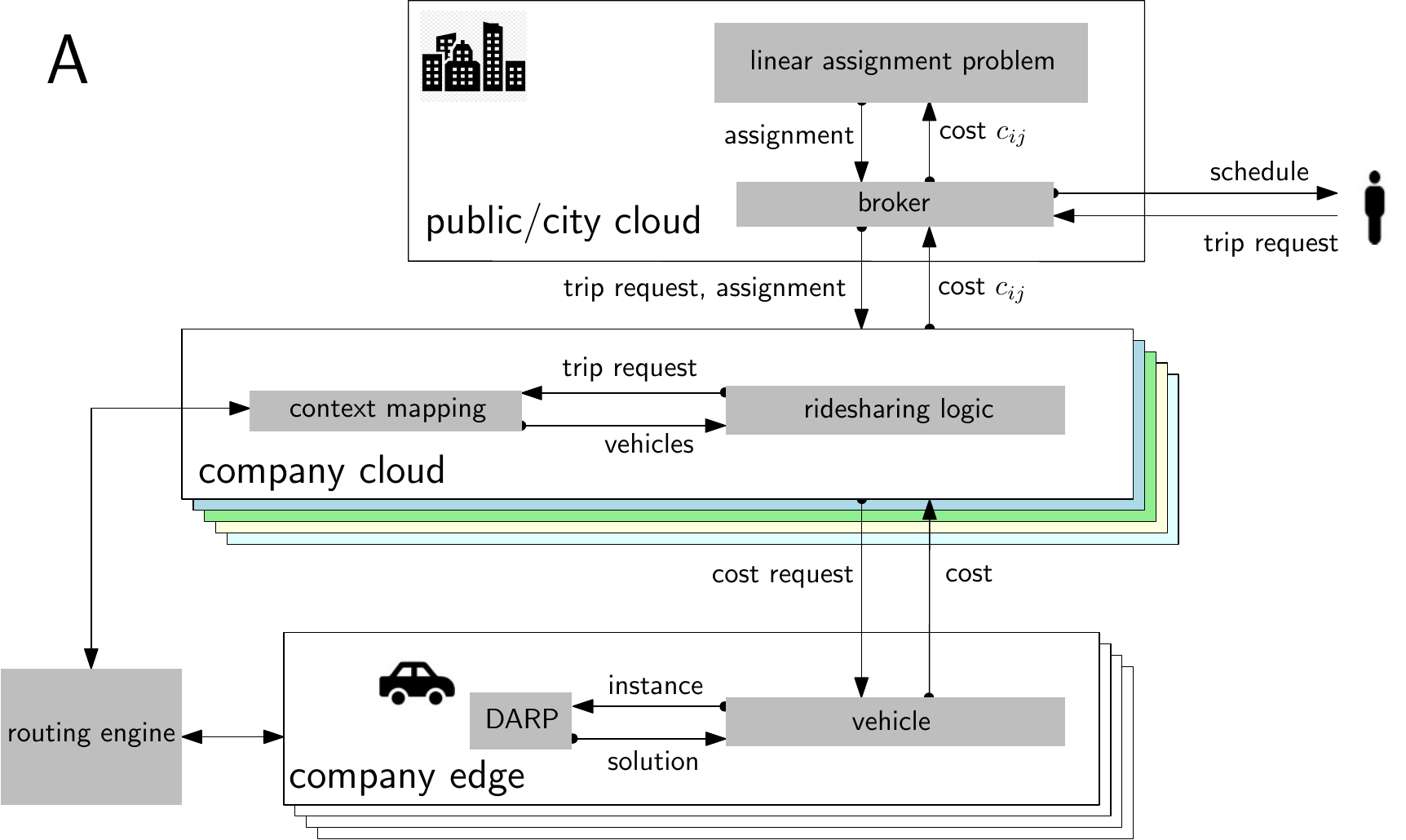}
	\caption{\Centr~architecture.}
	\label{fig:architectureA}
\end{figure}

As discussed, the need for centralization arises from the collection and solution of LAPs. We present next two different ways to overcome this point which yields cooperative and competitive solutions. 

In Figure~\ref{fig:architectureB}, we present the \Coop~architecture. We explicitly exploit the distributed auction algorithm of~\cite{Naparstek2014} to solve the LAP. In this architecture, the broker collects and sends customer requests and it receives bids from the companies and by successive iterations of the distributed algorithm, it can deliver the same optimal solution as if all the costs $c_{ij}$ were collected and the centralized \textbf{LAP(}$\mathcal{C},\mathcal{M}$\textbf{)} was solved. We will elaborate on these claims in Section~\ref{sec:coop}. The routing engine is the service that helps compute the costs needed for the DARP and assignment problems. It is important to note here that the initial LAP and the rebalancing one are solved iteratively (for each batch) via a distributed algorithm which collects the required information from the companies. In particular, the bids consisting of the difference between the 2 lowest costs for vehicles to serve each customer are collected by the broker. Customers are then assigned to vehicles with the highest bids, independently of the company. The process is repeated until all vehicles are assigned. The top-layer, i.e. the broker/city authority, coordinates this bidding process.  Note finally that as we shall see in Section~\ref{sec:coop} the number of iterations to reach optimality is bounded and, if latencies are handled correctly, one can expect this bidding process to be much faster than the sampling period.

\begin{figure}[H]
	\centering
	\includegraphics[width=0.6\textwidth]{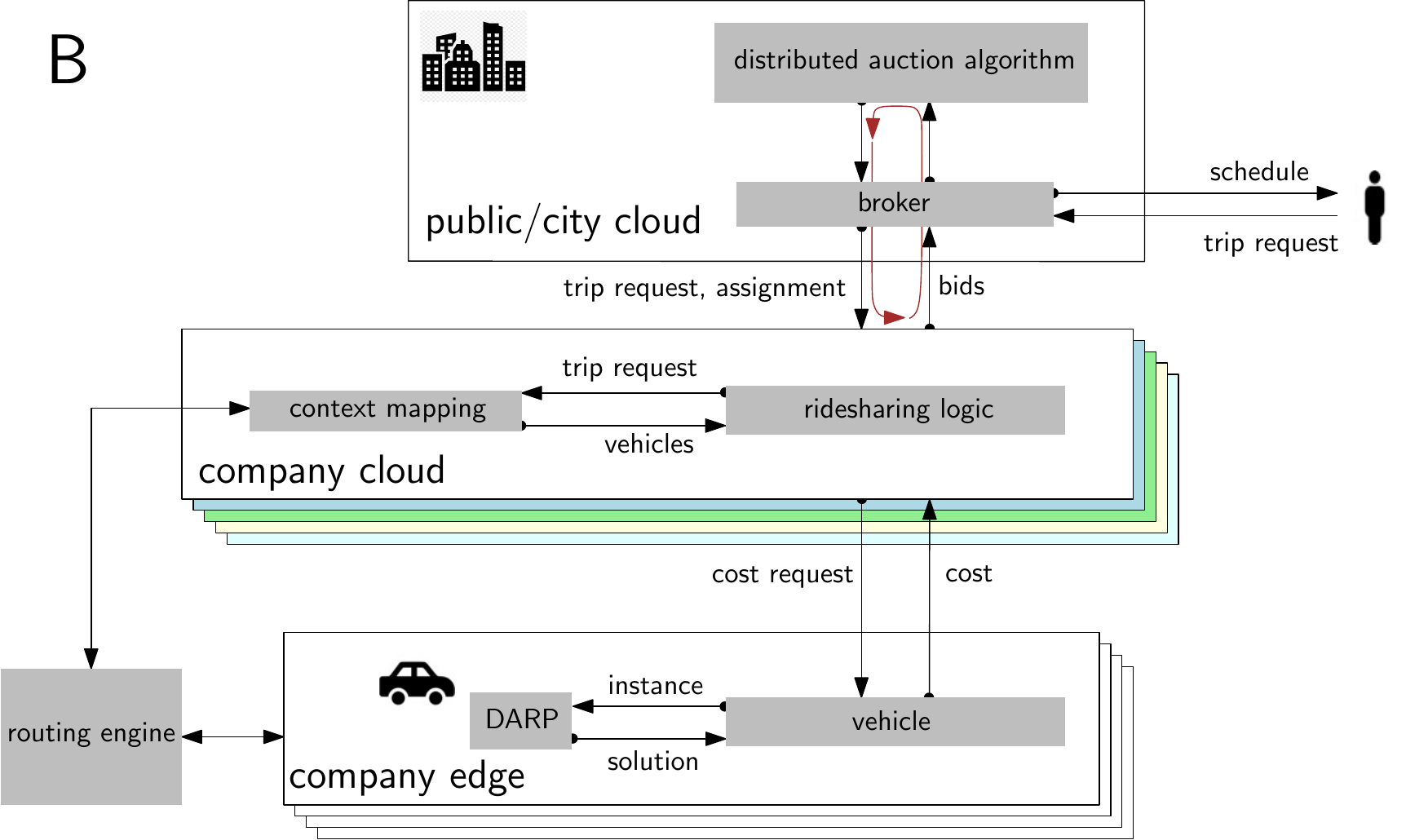}
	\caption{\Coop~architecture.}
	\label{fig:architectureB}
\end{figure}

In Figure~\ref{fig:architectureC}, we present the \Comp~architecture. In this architecture, the companies $p\in\mathcal{P}$ do not share costs or bids, but only desired customers to be served. In fact, the companies receive the trip requests via the broker and decide internally how to best accommodate them, by solving their own assignment problems \textbf{LAP(}$\mathcal{C}_p,\mathcal{M}$\textbf{)}. These LAP yield wanted one-to-one assignments that are then shared with the public cloud. Here the broker selects the best assignments for all the customers in their best interest, i.e. it select the lowest costs. The companies then re-iterate the competition on the unassigned customers, until all the customers that can be assigned, i.e., whose trip constraints are met, are assigned. The same process can be applied to rebalancing. 
  
This architecture may be slightly modified to accommodate customer preferences, which are known to affect the use of shared-mobility resources (see, e.g., \cite{KRUEGER2016343}) . Indeed, with customers preferences, the requests may be directly sent to the companies as opposed to the web-service, which corresponds to the situation in which customers directly book via the smartphone app of the ridesharing company. 

There is an important difference in the role of the broker between the \texttt{Cooperative} and \texttt{Competitive} architecture. In the \texttt{Cooperative} architecture, it coordinates and computes quantities. In the \texttt{Competitive} architecture, it is just collecting the assignments and selecting the best assignment, but that could also be done by the customers selecting the company offering the best deal.

\begin{figure}[H]
	\centering
	\includegraphics[width=0.6\textwidth]{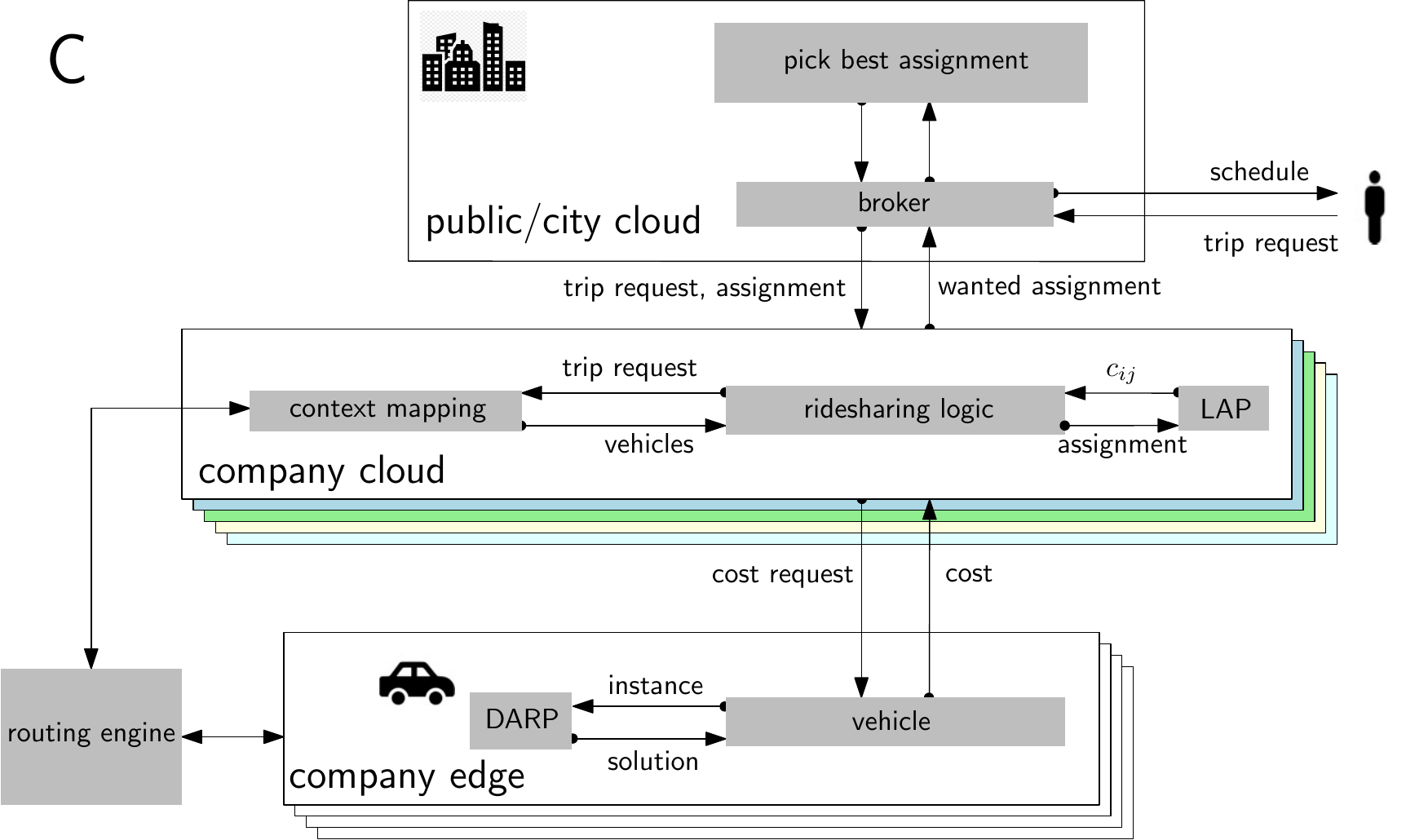}
	\caption{\Comp~architecture.}
	\label{fig:architectureC}
\end{figure}

We are now ready to present the details of the solution approach for both the \Coop~and \Comp~models. We have seen that the solution approach for the \Coop~model relies on the auction algorithm, particularly its distributed version~\citep{zavlanos2008distributed,Naparstek2014}, which we introduce in the Section \ref{sec:coop} \revision{(see~\cite{Chopra2017} for a distributed version of the Hungarian algorithm, which could be alternatively used here)}.

\revision{
	\paragraph{A note with respect to mechanism design}
	
	As one can infer by the proposed models, we are not considering here cases in which the companies are lying about what they can provide (in terms of travel times and constraint satisfaction) to achieve personal gain by gaming the system in their favor. Our models are, in general, not incentive compatible at a given time $t$, in the sense that companies have no incentives in telling the truth at a given time. Readers interested in mechanism that would enforce truth telling are referred to the vast literature in Vickrey-Clarke-Grove mechanisms and game theory, e.g.~\cite{Nisan2007}. 
	
	In this paper, we consider the realistic perspective that, if the companies would lie about what they could offer at time $t$, soon their customers would realize that change their preferences and go to another company at time $t'$. This is realistic, especially in very competitive settings like dynamic ridesharing in big cities with many companies, which are driven by customer satisfaction. In this context, companies have the incentive to tell the truth for competitive reasons in the long run (i.e., to keep customers) and truth-telling is a weakly-dominant strategy, that is: one is better off by being truthful, regardless of what the others do. 
	
	It is also important to point out that the main scope of the paper is to show that multi-company ridesharing could jeopardize the benefits of ridesharing (less vehicle on the road, better service), \emph{even when all the companies are telling the truth, and following customers' preferences}. We leave for future research the analysis of incentive-compatible mechanisms in the short and long time horizon.  
	
}
	
		

		

\section{\Coop~model}	
\label{sec:coop}

In this model, each company interacts with the broker which does not require access to the full $c_{ij}$ matrix of the costs. The broker orchestrates the cooperation between companies via a distributed auction algorithm, which is an adaptation of the implementation presented in~\cite{Naparstek2014}. As in the classical auction algorithm (\cite{bertsekas1988auction}), the bidding is the core concept of our implementation. In the distributed implementation, the players of the auction (i.e., the companies) send to the broker only the information needed to compute the bids $B(i,j)$ of each proprietary vehicle $i$ on the customer request $j$. Winning an iteration of the auction means having a vehicle assigned to a customer. The unassigned vehicles would raise their bids in the next iteration, resulting in a lower profit for the company. After a finite number of steps, \revision{the vehicles reach} a condition of \enquote{almost-equilibrium}, where there is no incentive for any vehicle to raise its bids. This condition corresponds to a feasible assignment of vehicles to requests, which is communicated to the companies by the broker.
In other words, the \Coop~ protocol provides an assignment solution where no company has any incentive to unilaterally seek another customer (by bidding higher). The almost-equilibrium condition is formally expressed by the so-called local $\epsilon-$complementary slackness (\lecs) condition 
\begin{equation}
-c_{ij_i} - B(i,j_i) \geq \max_{j\in\mathcal{M}} (-c_{ij} - B(i,j) ) - \epsilon \qquad \forall i \in \mathcal{C},
\label{eq:lecs}
\end{equation}
which ensures that each vehicle $i$ serves a customer $j_i$ that is within $\epsilon$ of the best current estimate of bid prices. Note that, in this context, profits are expressed as the assignment costs with opposite sign. Differently to the classical implementation of \cite{bertsekas1988auction}, the distributed auction of \cite{Naparstek2014} does not require the vehicles to disclose their best prices with the central broker, but their bids only. Specifically, in order to compute the bids at each iteration, the broker needs to receive the difference between the two best rewards for the unassigned vehicles, and not the entire assignment cost matrix (as a centralized protocol would require). 

Algorithm \ref{algo:disAlgo} describes the procedure for solving the assignment in the \Coop~protocol. At the beginning, all vehicles that are available $v \in \mathcal{C}_{\textrm{a}}$ are considered unassigned and the initial vehicle bids are set to zero. 
At each iteration, locally, each unassigned vehicle $i \in \mathcal{C}'$ computes the difference $\theta_i=\gamma_i-\omega_i$ between the two best possible net rewards obtained by serving (i.e., being assigned to) a customer. To participate to the auction, the unassigned vehicle $i$ raises its bid by $\theta_i$ and the perturbation $\epsilon$. Vehicles assigned to customers in previous iterations do not need to update their bid. The bid prices are collected by each company and sent to the broker, which assigns each customer to the highest vehicle bidder. The auction continues until all vehicles of each company are assigned, or a maximum number of iterations is reached. As mentioned in Section \ref{sec:prel}, the assignment returned by the auction algorithm needs to be interpreted for feasibility: vehicles assigned to requests with infinite costs (or in our implementation $c_{ij} = \frak{M}$, see Section \ref{sec:prel}) will not add the associated customers into their routes.

\paragraph{\rev{Example.}} \rev{To illustrate this, consider the simple scenario with two companies competing for two customers. The companies have one vehicle each. The two companies at the same time bid on the best customer for their vehicles, and send to the broker their favorite customer along with the difference in gain they would have if the second favorite customer was assigned to them. The broker updates the bid price for the customer and tentatively assigns the customers to the companies. The companies then can continue to bid, till no company has any incentive to unilaterally seek another customer by bidding higher (or the maximum number of iterations is reached), the final assignment is then enforced, and the algorithm terminates.  }

	
\begin{algorithm}[h]
\caption{\Coop~protocol for assignment of vehicles to customers}
\label{algo:disAlgo}
\begin{algorithmic}[1] 
\State Select $\epsilon >0$, and select a maximum number of iterations $k_{\text{coop}}$.
\State Set $k=0$, set all available vehicles $v\in \mathcal{C}_{\textrm{a}}$ as unassigned $\mathcal{C}' = \mathcal{C}_{\textrm{a}}$, and set bids $B(i,j)=0$, $\forall i,j$.
\While {There are unassigned customers that can be assigned and available vehicles \textbf{and} $k \leq k_{\text{coop}}$}
		\For {Each unassigned vehicle $i \in \mathcal{C}'$}
			\State Determine the maximum net reward $\gamma_i$ and the associated best customer $j_i$:
			$$
			\gamma_i= \max_{j \in \mathcal{M}} (-c_{ij}-B(i,j)), \quad j_i = \text{argmax}_{j \in \mathcal{M}} (-c_{ij}-B(i,j)).
			$$ 
			\State Determine the second  maximum net reward  $\omega_i$
			$$
			\omega_i = \max_{j \in \mathcal{M}, j \neq j_i} (-c_{ij}-B(i,j)).
			$$
		\EndFor
	 \For {Each company $p$}
					\For {Each unassigned  vehicle $i_p$}
						\State Send $(j_i, \gamma_i- \omega_i)$ to the broker.
				\EndFor
				\EndFor
		\State The broker updates bid price of each unassigned vehicle $i$ to its best customer $j_i$:
			\begin{equation*}
			B(i,j_i) = B(i,j_i) + \gamma_i-\omega_i + \epsilon.
			\end{equation*}
			\State Assigned vehicles bid on the last customer they bid on with the same price.
		\State  The broker assigns customers to the highest vehicle bidder $i^*$, and sends the assignment solution to the companies, i.e., $(j, j_{i^*})$
		\State Each company updates the assignment state of its vehicles and the set $\mathcal{C}'$ and the customer which they are assigned to.
				 \State Update iteration count $k \leftarrow k+1$.
			\EndWhile
			\end{algorithmic}
		\end{algorithm}
		

		
We next recall convergence properties of Algorithm~\ref{algo:disAlgo}. Proofs are omitted, given that they follow by those provided by \cite{Naparstek2014}. Theorem \ref{theo:conv} ensures that the distributed auction algorithm terminates in a finite number of steps and quantifies the deviation of the solution to optimality.
	
\begin{theorem}
	\label{theo:conv}
	The \Coop~algorithm converges to a feasible solution in a finite number of iterations. For a total fleet of $n$ vehicles, the solution is within $n\epsilon$ of the being the optimal assignment solution. If the assignment costs $c_{ij}$ are integer and $\epsilon < \frac{1}{m}$, then Algorithm~\ref{algo:disAlgo} terminates with the optimal assignment, as in the centralized case. 
\end{theorem}
	
For the practical implementation of Algorithm \ref{algo:disAlgo}, a primary question is the rate of convergence. Theorem~\ref{theo:bound} shows that the communication and computational complexity of the algorithm is polynomial in fleet size and the highest cost magnitude. This yields an upper bound on the communication overhead between companies and broker.


\begin{theorem}
	\label{theo:bound}
	Let $m$ be the number of requests, $n$ the number of vehicles, $C = \max_{i \in \mathcal{C}, j\in \mathcal{M}}|c_{ij}|>0$. Then the worst case maximum number of iterations for Algorithm~\ref{algo:disAlgo} to converge is $n m(1+C/ \epsilon)$. The worst case complexity of the \Coop~model is instead $O \left(n m^2(1+C/ \epsilon)\right)$.
\end{theorem}
	
\begin{proof}
Let $T_i$ be the number of iterations in which vehicle $i$ is not assigned, and $T$ the total number of iterations. First, note that the problem of maximizing $\sum -c_{ij}$ is equivalent to that of maximizing $\sum C-c_{ij}$, where $C=\max c_{ij}$. Since the auction matrix $(C-c_{ij})_{i \in \mathcal{C}, j \in \mathcal{M}}$ is non-negative, from Lemma $3$ by \cite{Naparstek2014}, 
\begin{equation}
T \leq \sum_{i\in \mathcal{C}} T_i \leq \sum_{i\in \mathcal{C}}\Big( m + m C/\epsilon - \sum_{j\in \mathcal{M}} c_{ij}/\epsilon\Big) \leq  n m (1+C/ \epsilon) \label{eq:boundT}.
\end{equation}
Since each iteration takes $O(m)$ computations (the computation of max for a vector of length $m$ is $O(m)$), then the computational complexity of the algorithm is $O \left( nm^2(1 + C/{\epsilon}) \right)$.
	\end{proof}
	
The communication and computational complexity is polynomial in the number of customers (note that $n=m$) and dependent on the magnitude of the perturbation factor $\epsilon$.

		

\section{\Comp~model}	\label{sec:comp}

In this model, the companies and the customers interact with a simpler broker logic. The companies solve a company-wide LAP to determine their best one-to-one assignments, which are then collected by the broker, which picks the best assignments among all, in an iterative way. The company-wide assignments are performed over a set $\mathcal{M}_p$ of available customers for company $p$. Initially, $\mathcal{M}_p=\mathcal{M}$ for all $p \in P$. Once the LAP is solved, the wanted assignments along with their costs are posted on the broker. 

If for a given customer $j \in \mathcal{M}$ multiple assignments are available (coming from different companies), the broker 
  selects the company 
    with the lowest cost and the corresponding vehicle/customer pair is considered assigned. This pair is then removed from the available vehicle and unassigned customer sets.  Each company continues to search for optimal assignment over the revised sets, until all customers are assigned. The algorithm terminates when all customers that can be assigned (i.e., all the ones for which all their time constraints are feasible) are assigned. 
    
\paragraph{\rev{Example.}} To illustrate this, consider the simple scenario with two companies competing for two customers. The companies have one vehicle each. The first company solves its LAP and assigns its vehicle to customer $1$ with cost $c_{11}$; the second company also assigns its vehicle to customer $1$ with cost $c_{21}<c_{11}$. The broker assigns then customer $1$ to the second company, and leave customer $2$ unassigned. In the next iteration, the first company chooses customer $2$ for its vehicle and the algorithm terminates.    

Algorithm~\ref{algo:noCommAlgo} shows the architecture of the algorithm. 

	\begin{algorithm}[h]
			\caption{\Comp~protocol for assignment}
			\label{algo:noCommAlgo}
			\begin{algorithmic}[1]
			\State Select a maximum number of iterations $k_{\text{comp}}$. 
			\State Set $k=0$, set all the available vehicles $v \in \mathcal{C}_{\textrm{a}}$ and customers as unassigned and set $\mathcal{M}_p= \mathcal{M}$, $\forall p, \quad \mathcal{C}' = \mathcal{C}_{\textrm{a}}$
			\While {there are unassigned customers that can be assigned and available vehicles \textbf{and} $k \leq k_{\text{comp}}$}
				\State Each company $p$ solves optimal assignment considering all its still unassigned vehicles $\mathcal{C}'$ (at current batch) and all customers in $\mathcal{M}_p$, i.e., \textbf{LAP(}$\mathcal{C}_p \cap \mathcal{C}', \mathcal{M}_p$\textbf{)}.
\For {Each company $p$}
	\State Send wanted assignment list to the broker.
\EndFor
\State The broker receives the assignments with their costs. The broker selects the company/vehicle that offers the lowest cost assignment (ties are broken randomly). 
\State Assigned customers and vehicles are removed from the still-to-assign sets and $\mathcal{M}_p$ and $\mathcal{C}'$ are updated.
\State Update iteration count $k \leftarrow k+1$.
			\EndWhile
			\end{algorithmic}
		\end{algorithm}
		
We now present a few properties of the algorithm. First, we find an upper bound on the number of iterations that needed for the algorithm to terminate (since the number of assignable customers is finite and at each iteration we assign at least one customer, finite termination is ensured). Second, we look at its worst case performance and derive a bound on the optimality gap between this algorithm and the centralized auction. Both properties are interesting: the first provides a tight communication overhead bound on the interaction between companies and the broker, the second provides an insight on the suboptimality of the competitive solution.

\begin{theorem}
\label{prop:NoCommMaxItr}
The maximum number of iterations $k_{comp}$ for termination of the \Comp~protocol with $P$ companies and $m$ new customers is upper bounded as $ k_{comp} \leq \log m/ \log [P/(P-1)] $. 
\end{theorem}
	
\begin{proof}
We assume the number of vehicles $n$ to be equal to the number of customers $m$; this is actually a worst case scenario, since at each iteration an equal number of customers/vehicles will be removed and if $n$ and $m$ are not equal, then $|n-m|$ vehicles or customers will not be assigned and the algorithm will terminate earlier. 

Let now $m^k$ denote the number of unassigned customers in iteration $k$. As per the algorithm initialization, $m^0 = m$. Let $n_p^k$ denote the number of unassigned vehicles for company $p$ at the beginning of iteration $k$.
		
At each iteration $k$, each company $p$ solves its assignment problem and sends its wanted assignments. At the end of iteration $k$, at least $\bar{m}$ new customers will be assigned, where $\bar{m} = \max_{p \in P} n_p^k$, i.e., we assign at least a number of customers equal to the number of vehicles of the biggest company (since in the worst case all the other companies will compete for the same customers). Therefore, the maximum number of unassigned customers in iteration $k+1$, is given by $m^{k+1} = m^k -  \max_{p \in P} n_p^k$. We now put ourselves in the scenario in which all the companies have an equal number of vehicles; this is again a worst case scenario for $m^{k+1}$, and in particular we can write
\begin{equation}
m^{k+1} = m^k - n_p^k = m^k - \lfloor m^k/P \rfloor = \lfloor m^k (P-1)/P\rfloor \leq m^k (P-1)/P = m^0 ((P-1)/P)^k,
\end{equation}
where $P$ is the number of companies, and $\lfloor \cdot \rfloor$ is the floor operator. The number of iterations for $m^{k+1} = 0$ is the one for which $m^0 ((P-1)/P)^k <1$, so that $\lfloor m^k (P-1)/P\rfloor = 0$, and solving for $k$, $k > \log m^0/ \log [P/(P-1)]$. Therefore a bound $k_{comp}$ for the maximum number of iterations, such that if $k  > k_{comp}$ the algorithm terminates,  is the right-hand side of the previous inequality and the proposition is proven. 
\end{proof}

\Cref{prop:NoCommMaxItr}} states that the number of iterations goes logarithmically with the number of requests $m$. 

We then move on to proving a worst-case bound for the competitive algorithm. In contrast to the distributed assignment case, the competitive case is not guaranteed to converge to optimal. Rather, we next show that in the worst scenario, the competitive case can lead to solutions which are at least three time \rev{worse than} the optimal solution. For this theorem, we will need a reasonable additional assumption: the costs need to satisfy a triangle-like inequality: given the costs $c_{ab}$, $c_{ac}$, $c_{dc}$, $c_{db}$, then we require that $c_{ab} \leq c_{ac} + c_{dc} + c_{db}$; that is the cost of vehicle $a$ moving from its position to the position of $b$ is less than the cost of moving from $a$ to $c$ plus moving from $d$ to $c$ and $d$ to $b$.   

In our case the costs represent travel times, so the assumption is reasonable assuming that travel time from point $x$ to $y$ are close enough to travel time from $y$ to $x$ (in this case the inequality just say that the travel time from point $a$ to $b$ is not longer than the sum of the travel times between $a$ and $b$ passing through position $c$ and $d$). 
	
\begin{theorem}
\label{theo:noComm}
Assume that the costs satisfy a triangle-like inequality: given the costs $c_{ab}$, $c_{ac}$, $c_{dc}$, $c_{db}$, then $c_{ab} \leq c_{ac} + c_{dc} + c_{db}$. Under this assumption, the following upper bounds hold. 
		
For two companies, the competitive algorithm's total cost in the worst case is two times as much as the one of the best cooperating solution. For three companies, it is at worst three times as much. For an higher number of companies, the worst case is at least three times as much.
\end{theorem}

\begin{proof}
The proof is constructive and finds the worst possible scenario. We will proceed in the following way, \emph{(i)} we argue that worst case is the one with $n$ vehicles and $n$ requests; \emph{(ii)} we argue that the worst case is the one with $n$ companies with $1$ vehicle each; \emph{(iii)} we find the worst possible case for $n=2$ by linear programming and show that the competitive protocol delivers a solution at most twice as much as the one of the best cooperating solution; \emph{(iv)} we generalize the $n=2$ case for $n=3$ in the same way; finally \emph{(v)} we generalize for $n>3$. 
	
	Point \emph{(i)}. Imagine that there are $n$ vehicles and $m$ requests; if $n>m$ then only $m$ vehicles can be assigned. Indicate with $\mathcal{S}_{c}$ the set of assigned vehicles for the cooperating solution and $\mathcal{S}_{nc}$ the set of assigned vehicles for the competitive solution. If  $\mathcal{S}_{c}=\mathcal{S}_{nc}$, we can safely remove $n-m$ vehicles and nothing will change. If  $\mathcal{S}_{c}\neq \mathcal{S}_{nc}$, it means that in the competitive scenario, selecting the vehicles in $\mathcal{S}_{c}$ would incur in a higher cost, therefore sticking with the $m$ vehicles in $\mathcal{S}_{c}$ is the worst case.  We can proceed analogously for the case $n<m$ and conclude that the case $n=m$ is the worst possible case. 
	
Point \emph{(ii)}. Imagine that a company has multiple vehicles, then the company would chose the bidding optimally within its fleet and avoid conflicts. This lowers the total cost. Therefore the worst case is the one when each company is competing with each other one, and they have $1$ vehicle each. 
	
\begin{figure}[ht!]
\centering
\includegraphics[width=.5\columnwidth]{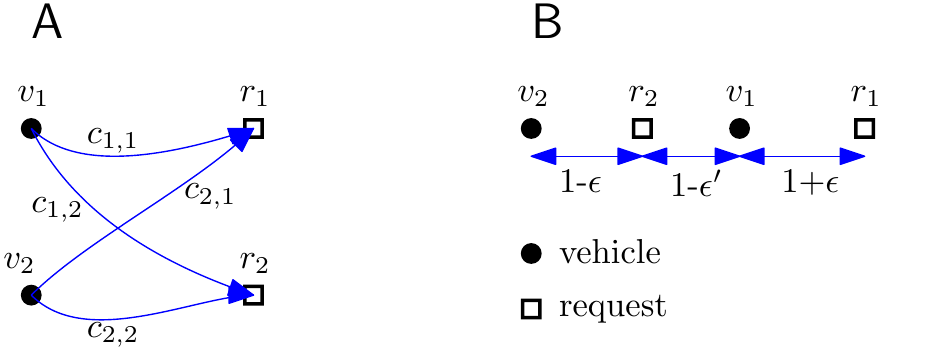}
\caption{Proof figure for the $n=2$ case.}
\label{fig:NoCommProof-1}
\end{figure}
	
Point \emph{(iii)}. Consider Figure~\ref{fig:NoCommProof-1}.A, where circles represent vehicles and squares represent requests. Here we indicate costs with the notation $c_{i,j}$ to avoid labelling confusion. Let the best cooperating solution be $(v_1,r_1)$ and $(v_2,r_2)$, meaning that we are selecting the costs $c_{1,1}$ and $c_{2,2}$, and the total cost is $c^* = c_{1,1}+c_{2,2}$. Set the scale without loss of generality to $2$, so that $c^* = 2$. The worst possible case is that, in the competitive protocol, we switch the solution to $(v_1,r_2)$ and $(v_2,r_1)$, so that $c = c_{1,2}+c_{2,1}$. For this to happen, either both $v_1$ and $v_2$ have to select $r_1$ or $r_2$, and then one of them has to ``settle'' for the less appealing request. With no loss of generality, we choose that $v_1$ selects $r_2$ with highest bid, while $v_2$ selects $r_2$, but then has to settle for $r_1$. For this case, we need that $c_{1,2}< c_{1,1}$, $c_{2,2}<c_{2,1}$; for optimality of the cooperating solution, $c_{1,1}+c_{2,2}< c_{1,2}+c_{2,1}$, while for triangle inequality, $c_{2,1}\leq c_{2,2}+c_{1,2}+c_{1,1}$. Putting all these relations together, we want to maximize $c = c_{1,2}+c_{2,1}$ subject to some linear equalities and inequalities, i.e., we want to solve the linear program:
	\begin{align*}
	\max_{c_{1,1}, c_{1,2}, c_{2,1}, c_{2,2}} & c_{1,2}+c_{2,1} \\
	\textrm{subject to } & c_{1,1}+c_{2,2} = 2, \\
	& c_{1,2}< c_{1,1}, \quad c_{2,2}<c_{2,1}, \\
	& c_{1,1}+c_{2,2}< c_{1,2}+c_{2,1}, \\
	& c_{2,1}\leq c_{2,2}+c_{1,2}+c_{1,1}, \\
	& c_{1,1}, c_{1,2}, c_{2,1}, c_{2,2}\geq0. 
	\end{align*}
The solution of the linear program is $c_{1,1}=1+\epsilon_1, c_{1,2}=1-\epsilon_2, c_{2,1}=3-\epsilon_2, c_{2,2}=1-\epsilon_1$, with $\epsilon_1>\epsilon_2>0$ very small positive constants, and $c = 4-2\epsilon'$. Therefore $c/c^* < 2$ and the competitive protocol delivers a solution at most twice as much as the one of the best cooperating solution (shown in Figure~\ref{fig:NoCommProof-1}.B).
	
Point \emph{(iv)}. We proceed for the case $n=3$ in the same way that in the case $n=2$, by constructing worst cases linear programs. Let the best cost be $c^* = \sum_{i=1}^3 c_{i,i} = 3$. The worst cost is achieved when all the vehicles pick up different customers, and we can fix it w.l.g. at $c = \sum_{i=1}^2 c_{i,i+1} + c_{3,1}$. To reach such configuration, we can either have two conflicts (e.g., all of the vehicles selecting customer $2$, then vehicles $2,3$ both selecting customer $3$), or one conflict (e.g., vehicle $1$ selecting customer $2$, and vehicles $2,3$ both selecting customer $3$). In both cases, we can write the resulting linear program and obtain $c<9$. Therefore $c/c^* <3$. 

Point \emph{(v)}. We could proceed as in the cases $n=2,3$ for the cases $n>3$, yet the number of possible cases increase and therefore it becomes more complicated to construct all the possible linear programs. In general however, one can say that in the \emph{worst case}, if $c^* = n$, then $c$ is at least $3n$. 
\end{proof}

\Cref{theo:noComm} indicates that the solution provided by competing companies may be significantly sub-optimal at each time with respect to the best cooperating solution. This is worrisome in the context of dynamic ridesharing; however, we will show in the simulations in \Cref{sec:num-real} that, although the solution for each of the LAP is sub-optimal, in the long run (that is considering dynamic scenarios and a week long simulation) the performance of the competitive algorithm is not so poor. This is an important point, also made in~\cite{simo2018demand}: optimality at a given time is not a proxy for optimality in the long run. 

		

\section{Numerical studies}	\label{sec:num}

In this section, we present the numerical studies to test our algorithms and model. In \Cref{sec:num-static}, we conduct a sensitity analysis present results from a static analysis to showcase the performance of the different algorithms in solving single-periods ridesharing models. Then, in \Cref{sec:num-real} we report realistic 1-week simulations, using the New York City data set. Finally, \Cref{sec:num-findings}  describes the main findings obtained by the simulations.

\subsection{Sensitivity analysis of the ridesharing models}\label{sec:num-static}

We aim to compare the performance of the \Centr~model with the \Coop~and \Comp~models and their variants. We conduct tests on three types of randomly generated cost matrices with 100, 500, and 1000 customers. The cost matrix elements are \revision{sampled from the travel time values between different locations from the New York City taxi public dataset~\citep{NYCdata}}. Three other control variables are regulated in the experiments including the number of companies, the vehicle-to-company mapping, and the proportion of vehicles belonging to each company, i.e. the market share. \revision{The following results are reported as an average over 20 random instances of each cost matrix.}


First we report analysis for the \Coop~model. We compare the percent cost difference from the optimal, i.e. optimality gap, if the cost matrix is stochastic or has a percent bias for each player. A true cost matrix is first randomly sampled as discussed earlier. Then, a stochastic noise is added on to it following a Gaussian distribution with the mean as the true cost matrix value and the chosen value of standard deviation. Then, a bias term is added to the cost value. The bias may be positive or negative. Three levels of standard deviation are considered: no standard deviation, low standard deviation of \revision{1 minute} (\texttt{LowSD}), and high standard deviation of 2 minutes (\texttt{HighSD}). Three levels of bias are considered: no bias, low bias ranging randomly between 0-10\%, and high bias ranging randomly between 40-50\%. The bias term may be randomly positive or negative. The perturbation $\epsilon$ is set to 0.01 (see Algorithm~\ref{algo:disAlgo}). Table \ref{tab:coopStochBiasStatic} shows the average optimality gap in \% in the case of two companies with equal market share. The average is reported over 500 different instances of stochasticity and bias \revision{for the case of 100 customers and two companies with equal market share}.

\begin{table}[h]
	\scriptsize
	\centering
	\caption{Average optimality gap (\%) for the \texttt{Cooperative} model under the presence of stochasticity and bias.}
	\label{tab:coopStochBiasStatic}
	\begin{tabular}{|c|l|l|l|}
		\hline
		& \textbf{No bias} & \textbf{Low bias (0-10\%)} & \textbf{High bias (40-50\%)} \\ \hline
		\textbf{No stochasticity} & 0.00 & \revision{0.07}\% & \revision{1.72}\% \\ \hline
		\textbf{\texttt{LowSD}} & \revision{20.78}\% & \revision{21.93}\% & \revision{41.61}\% \\ \hline
		\textbf{\texttt{HighSD}} & \revision{55.88}\% & \revision{56.81}\% & \revision{98.16}\% \\ \hline
	\end{tabular}
\end{table}

As expected, the optimality gap increases with the standard deviation and the bias. It is interesting to note the high sensitivity to stochastic noise.

\begin{figure}[H]
	\includegraphics[width=0.5\columnwidth]{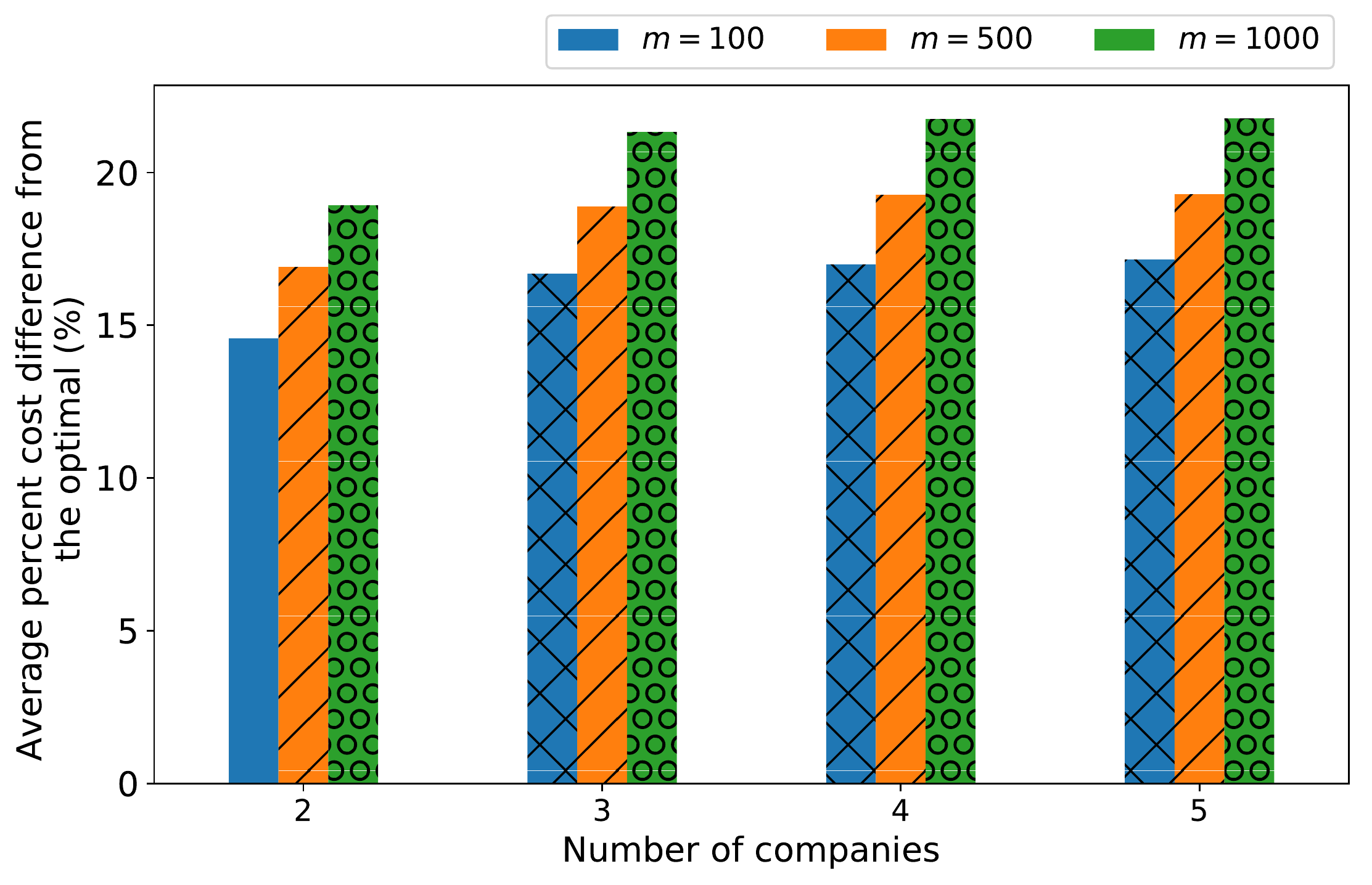}
	\includegraphics[width=0.5\columnwidth]{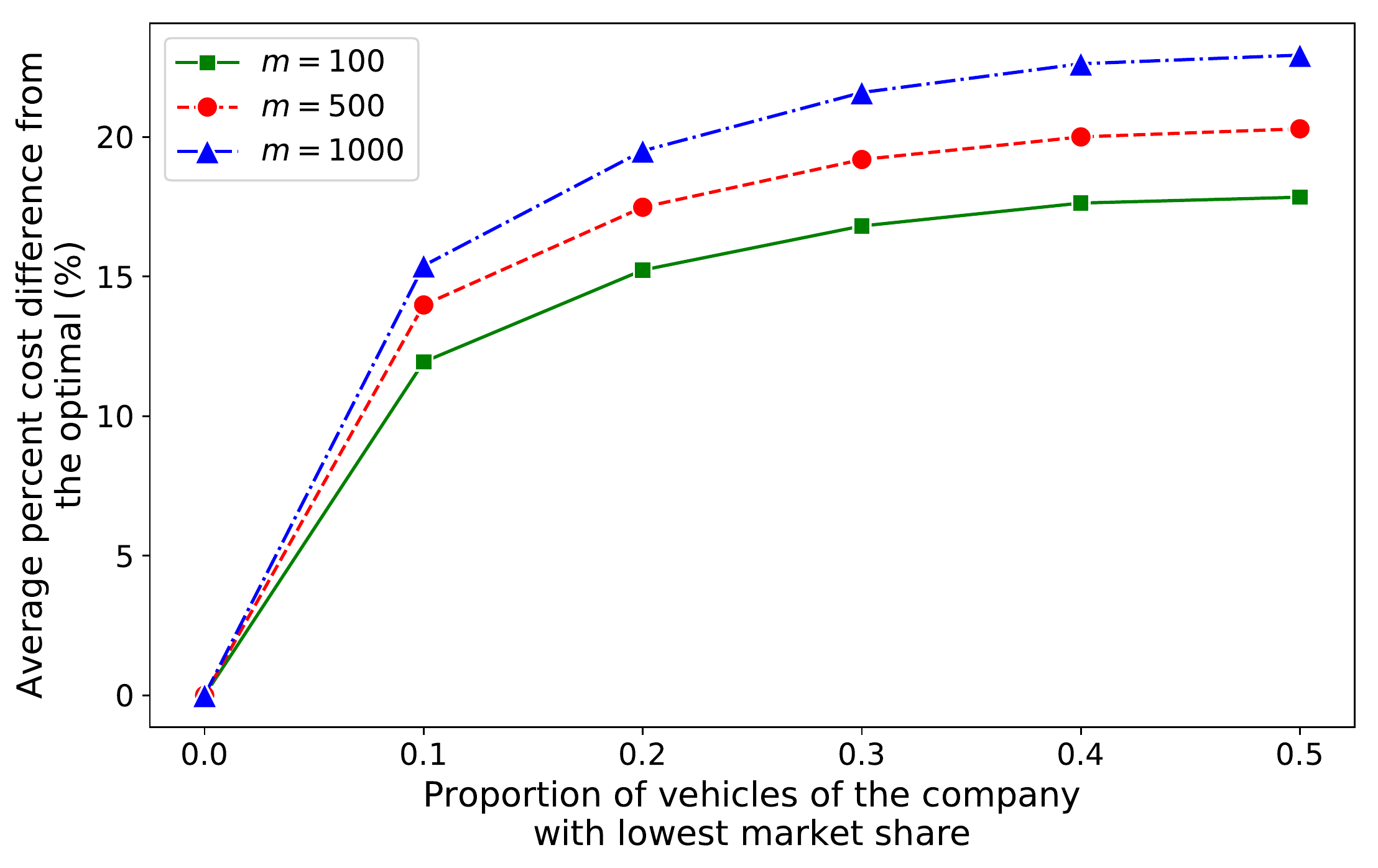}
	\caption{(a) Variation of average optimality gap for different sizes of cost matrices and for different number of companies, and (b) Variation of average optimality gap with varying market penetration of the company with smallest market share in the case of two companies.}
	\label{fig:competitiveStaticResult}
\end{figure}

We now report results for the \Comp~model. We consider a varying number of companies, respectively 2, 3, 4 and 5 companies. We generate 500 randomized instances of market settings with varying vehicle-to-company mapping and varying market share for each company. For each instance, the cost of \Comp~model is compared with the optimal costs and the average difference over the 500 instances is reported. \revision{Figure~\ref{fig:competitiveStaticResult}(a)} reports the optimality gaps for different number of customers, and different number of companies.

We make following observations. First, when the number of companies increase, it increases the optimality gap. The increase is the \revision{highest} when a third company is added: the optimality gap increases by an average \revision{2.17\%} after going from a two-company scenario to a three-company scenario. This confirms the intuition that more competition worsens the system performance. Second, the optimality gap \revision{also increases with the number of customers. This is also as expected where having more customers amplifies the impact of competition and leads to higher percent cost difference.}


We also evaluate the optimality gap for varying market share in the two-company scenario. Again, we generate 500 instances of the vehicle-to-company mapping for each given market share. \revision{Figure \ref{fig:competitiveStaticResult}(b)} shows the evolution of the optimality gap with respect to the market share of the company with lowest market share. We observe that the more skewed the market share is in favor of one company, the lower the optimality gap is, and, as expected, the costs are optimal when the market is monopolistic. It is interesting to note that the situation of equal market share leads to worse assignments.

We finally look at the evolution of the optimality gap in the presence of customers preferences. Each customer may have a preference for any company and may have a switching threshold. As stated in Section~\ref{3-model}, this means that if the travel time offered by the non-preferred company is lower than the travel time offered by the preferred company minus the given threshold, then the customer chooses to switch to the other company. We evaluate varying percentage of customers with preferences and varying values of switching threshold. We perform the tests for two-company scenario with equal market share and with 100 and 1000 customers. Again, the average percent cost difference is reported over 500 instances of vehicle-to-company mapping. Table~\ref{tab:compPrefStatic} shows the evolution of the optimality gap for different values of switching threshold and different percentage of customers with preferences. The cells with the highest optimality gap are highlighted.

\begin{table}[h]
	\scriptsize
	\centering
	\caption{Average percent optimality gap for the \Comp~model under the setting of two companies with equal market share with varying percentage of customers with preferences, and varying switching thresholds.}
	\label{tab:compPrefStatic}
	\begin{tabular}{|l|l|l|l|l|l|l|l|l|l|l|}
		\hline
		\multicolumn{2}{|l|}{}                                                                                                      & \multicolumn{4}{c|}{\textbf{100 riders}}                                                                                                                                                                                                                    &                & \multicolumn{4}{c|}{\textbf{1000 riders}}                                                                                                                                                                                                                   \\ \cline{3-11}
		\multicolumn{2}{|l|}{}                                                                                                      & \multicolumn{4}{c|}{Switching threshold}                                                                                                                                                                                                                    &                & \multicolumn{4}{c|}{Switching threshold}                                                                                                                                                                                                                    \\ \cline{3-11}
		\multicolumn{2}{|l|}{\multirow{-3}{*}{}}                                                                                    & \multicolumn{1}{c|}{\begin{tabular}[c]{@{}c@{}}0\\   (Always \\ switch)\end{tabular}} & \multicolumn{1}{c|}{1 min} & \multicolumn{1}{c|}{5 min}              & \multicolumn{1}{c|}{\begin{tabular}[c]{@{}c@{}}High \\ (never\\   switch)\end{tabular}} &                & \multicolumn{1}{c|}{\begin{tabular}[c]{@{}c@{}}0\\   (Always \\ switch)\end{tabular}} & \multicolumn{1}{c|}{1 min} & \multicolumn{1}{c|}{5 min}              & \multicolumn{1}{c|}{\begin{tabular}[c]{@{}c@{}}High \\ (never\\   switch)\end{tabular}} \\ \hline
		& \textbf{0\%}   & \revision{10.17\%}                                                                               & \revision{10.17\%}                                                                                                  & \revision{10.17\%}                                                                                                                & \revision{10.17\%}                                                                               & \textbf{0\%}   & \revision{25.42\%}                                                                               & \revision{25.42\%}                                                                               & \revision{25.42\%}                                                                                                                & \revision{25.42\%}                                                                               \\ \cline{2-11}
		& \textbf{20\%}  & \revision{10.17\%}                                                                               & \revision{14.19\%}                                                                                                 & \revision{13.48\%}                                                                                                                                  & \revision{14.14\%}                                                                                                                                 & \textbf{20\%}  & \revision{25.42\%}                                                                               & \revision{26.17\%}                    & \revision{26.20\%} & \revision{26.14\%}  \\ \cline{2-11}
		& \textbf{40\%}  & \revision{10.17\%}                                                                               & \revision{16.18\%}                                                                                                                   & \revision{16.71\%}                                                                                                                                                                  & \revision{16.24\%}                                                                                                                                 & \textbf{40\%}  & \revision{25.42\%}                                                                                                                                                                                                                                                 & \revision{26.79\%}                     & \revision{27.03\%} & \revision{26.61\%} \\ \cline{2-11}
		& \textbf{60\%}  & \revision{10.17\%}                                                                               & \revision{17.99\%}                                                                                                                   & \revision{18.67\%}                                                                                                                                                                   & \revision{17.36\%}                                                                                                                                                                                                                     & \textbf{60\%}  & \revision{25.42\%}                                                                               & \revision{27.28\%}                     & \revision{27.32\%} & \revision{27.26\%} \\ \cline{2-11}
		& \textbf{80\%}  & \revision{10.17\%}                                                                               & \revision{19.58\%}                                                                                                                   & \revision{18.99\%}                                                                                                                                                                   & \revision{19.75\%}                                                                                                                                 & \textbf{80\%}  & \revision{25.42\%}                                                                               & \revision{27.92\%}                    & \revision{27.70\%}                                  & \revision{28.16\%}                                                                                     \\ \cline{2-11}
		\multirow{-6}{*}{\textbf{\begin{tabular}[c]{@{}l@{}}Percent \\ riders \\ with \\ preference\end{tabular}}} & \textbf{100\%} & \revision{10.17\%}                                                                               & \revision{21.27\%}                                                                                                       &  \revision{20.89\%}                                                                                                                                             & \cellcolor[HTML]{6195C9}\revision{21.29\%}                                                                                                                                  & \textbf{100\%} & \revision{25.42\%}                                                                               & \revision{28.76\%}        & \revision{28.57\%}            & \cellcolor[HTML]{6195C9}\revision{28.78\%}  \\ \hline
	\end{tabular}
\end{table}

As expected, an increase of the percentage of customers with preferences and of the switching threshold deteriorates the quality of the assignments. It is interesting to note that the deterioration caused by the increase of the switching threshold is not that consequent. 

We shall corroborate these findings in the subsequent section, in the light of real-world data.

\subsection{Real-time dynamic simulations}\label{sec:num-real}

We now consider real-time on-demand ridesharing and the effect of cooperation and competition on the quality of the service provided.

We consider the New York City taxi public dataset~\citep{NYCdata} and extract one week of data trips from 00:00 hours on Sunday, May 5, 2013 to 23:59 hours on Saturday May 11, 2013. This dataset contains the time and location of all of the pick-ups and drop-off locations visited by the 13,586 active taxis, for each day. From these data, we extract the origin-destination requests, as well as the time of pickup (which is the time of request). The number of requests each day varies between 382,779 (Sunday) and 460,700 (Friday). We consider the complete road network of Manhattan as encoded in Open Street Map, amounting at 17,446 nodes.

The fleet is initialized in random locations within Manhattan at 22:00 hours on Saturday, May 4, 2013, (and the requests between 22:00 and 23:59 are used to warm start the fleet, so that the vehicles are not all empty at the beginning of the service, while the requests are batched and sent to the ridesharing service every $h = 10$ seconds. We consider that each trip has constraints on the maximum delay time (how much a customer is willing to wait for the ride) and detour time (how much additional trip time from the original schedule the customer is willing to accept to allow the vehicle to accommodate other customers) of $7$ minutes for both. In all the simulations, the cost function is the total trip duration. The ridesharing algorithms are implemented in Python 2.7 on a 2.7GHz Intel i5 laptop with 8GB RAM memory. 



In this section, we want to assess the \Coop~and \Comp~models for multi-company ridesharing. We consider three companies, with a number of vehicles proportional to the number of trips per day reported in~\cite{NYCreports}. In particular, company 1 has 265 vehicles, company 2 has 175, and company 3 has 60, for a total of 500 vehicles on the road. All the vehicles have 4 seats available. With this setting, as reported in~\cite{alonso2017demand,simo2018demand}, a centralized ridesharing service would be able to offer a very high quality of service to about 1/6 of the NYC current demand. Therefore, in our simulations we consider 1/6 of the total trip requests. Indeed, as explained in~\cite{simo2018demand}, considering the whole demand and a corresponding $3000$ vehicle fleet would lead to quantitatively similar results and the main conclusions that we report here will not change. The reason to use a smaller fleet size and corresponding demand is to keep the computational time manageable to run several representative scenarios. 

The number of vehicles considered in the context mapping module for each company is set to $10$. Specifically, it is assumed that each company has similar computational resources and attempts to allocate each request to a maximum of $10$ candidate vehicles within a given radius from the requests by solving $10$ DARPs. It is important to note that this is an upper bound, but in many cases (especially for small companies), even fewer vehicles will be inside the considered radius from each request. This choice in the context mapping seeks a trade-off between computational requirements and quality of the ridesharing service. It is reasonable to assume that the software and hardware requirements are similar for the comparnies. 
 All algorithmic variants include a rebalancing service described in~\cite{simo2018demand}, which the customers are accepting: \cite{Alonso-Mora2017} and \cite{simo2018demand} showcased the greatly positive impact of rebalancing in dynamic ride-sharing.
 
We measure performance with different metrics. Firstly, we consider the service rate (SR), defined as the ratio of serviced customers on the total number of customers. Clearly, the higher the SR is, the better the quality of a ridesharing service is. Secondly, we look at customers' waiting times and vehicles' detour times, both cumulatively and per company. Low values of both indicators imply a high level of service. However, a comparable value across the companies is usually desired as well, in order to limit the dissatisfaction of customers. Finally, we look at how the customers are assigned among the companies. If the assignment is proportional to the number of vehicles, we say that the algorithm reaches an egalitarian solution, otherwise an unbalanced high level of service of specific companies may be deemed as unfair.

In the simulations, we remark that searching myopically for optimality at a given time is not a proxy for optimality in the long run. In other words, suboptimal vehicle routes at a given time do not necessarily degrade the performance in the long run (e..g, after a week), because they might help to accommodate more requests in future time periods. We observed that this is true also for the \Coop~and \Comp~cases. This is an important concept in dynamic in real-time ridesharing, as pointed out and discussed by~\cite{simo2018demand} in the single-company setting.

We test and compare the \Coop~model (simulations 2-8) to the \Centr~model (simulation 1) in Table~\ref{tab:benchmark-coop}. We report: the number of iterations of the distributed auction algorithm for the \Coop~model, the stochasticity and bias added (which will be explained in detail in the next paragraph), the service rate obtained, the difference in percentage with a perfect egalitarian solution of customers served (which sums up to 0) and the total percentage of served customers (which sums up to the SR), the waiting time and detour time (cumulative and per company), the vehicle occupancy per company, and the average computational time for each assignment. As expected, the results of the centralized algorithm in row 1 are in-line with the ones presented in~\cite{simo2018demand}. In particular, note the level of service is greater than $99\%$, and the assignment solutions are close to the egalitarian share of $53\%, 35\%, 12\%$ customers assigned to the three companies, respectively. The discrepancy with a perfect egalitarian partition is due to the local context mapping, which restricts the number of available vehicles for each customer, and alleviates the computational burden. The computational time is well less than 10~\si{\second}, and therefore the algorithm can be implemented in real-time.

\setlength\tabcolsep{4.5pt}
\begin{table}
	\scriptsize
	\centering
	\caption{Simulations for the \Coop~scenario with three companies, having $265, 175, 60$ vehicles, respectively, which correspond to a 53\%, 35\%, 12\% share. The two rows for the \textit{customers} column represent the difference in percentage w.r.t. an egalitarian partition of the serviced customers and the total percentage of the serviced customers, respectively. Waiting and detour times are reported both in cumulative and per-company values.}
	\label{tab:benchmark-coop}
	\begin{tabular}{ccccccccccc}
		\toprule
		sim. & no. iters & stoch & bias  & SR & customers & waiting &  detour   &  occupancy & comp. time\\
		& & [min] & [\%] & [\%] & [\%] & [min]   &  [min]  &  [cust./vehicle] & [s] \\
		\toprule
		\rowcolor{Gray} 
		1 &  Centr. & 0,0,0 & 0,0,0 & 99.4 & \mul{-2.3,0.8,1.5}{50.4, 35.6, 13.4}& \mul{3.6}{3.6, 3.6, 3.6} & \mul{2.9}{2.8, 2.9, 3.2}  & 0.7, 0.8, 0.9 & 1.5\\
		\midrule
		2 & 1000 & 0,0,0 & 0,0,0 & 98.9 & \mul{-0.8,0.6,0.2}{51.6, 35.2, 12.1} & \mul{3.2}{3.2, 3.2, 3.4} & \mul{2.6}{2.6, 2.6, 2.9}  & 0.6, 0.7, 0.8 & 7.3\\
		\midrule
		\rowcolor{Gray}
		3  & 500 & 0,0,0 & 0,0,0 & 92.4 & \mul{-0.7,1.4,-0.7}{48.3, 33.6, 10.5} & \mul{3.2}{3.2, 3.2, 3.3} & \mul{2.6}{2.5, 2.6, 2.9}  & 0.6, 0.6, 0.6 & 5.8\\
		4  & 250 & 0,0,0 & 0,0,0 & 90.6 & \mul{-0.4,1.5,-1.1}{47.7, 33.1, 9.8} & \mul{3.2}{3.1, 3.2, 3.2} & \mul{2.4}{2.4, 2.4, 2.7}  & 0.5, 0.6, 0.6 & 4.0\\
		\midrule
		\rowcolor{Gray}
		5  & 1000 & 5,5,5 & 0,0,0 & 98.5 & \mul{-1.9,0.0,1.9}{50.3, 34.5, 13.7} & \mul{4.1}{4.0, 4.2, 4.5} & \mul{3.1}{3.0, 3.1, 3.4}  & 0.7, 0.7, 1.0 & 5.7\\
		6 & 1000 & 10,10,10 & 0,0,0 & 97.9 & \mul{-2.9,0.7,2.2}{49.1, 35.0, 13.8} & \mul{4.3}{4.2, 4.4, 4.6} & \mul{3.5}{3.4, 3.5, 3.8}  & 0.7, 0.8, 1.1 & 4.8\\
		\midrule
		\rowcolor{Gray}
		7  & 1000 & 0,0,0 & 0,0,10 & 98.9 & \mul{-2.2,0.6,1.6}{50.2, 35.2, 13.5} & \mul{3.2}{3.1, 3.2, 3.6} & \mul{2.7}{2.5, 2.6, 3.0}  & 0.6, 0.6, 1.0 & 7.1\\
		8  & 1000 & 0,0,0 & 0,0,20 & 98.9 & \mul{-2.5,0.3,2.2}{49.9, 34.9, 14.1} & \mul{3.2}{3.0, 3.1, 3.9} & \mul{2.6}{2.4, 2.5, 3.2}  & 0.6, 0.6, 1.2 & 7.0\\
		\toprule
	\end{tabular}
\end{table}

\paragraph{\texttt{Cooperative} model.} The performance assessment of the \Coop~model is presented in~\Cref{tab:benchmark-coop}:

\begin{enumerate}
\item Baseline (row 2): a distributed auction algorithm that implements the \Coop~model and uses 1000 iterations for each linear assignment problem. No stochasticity or bias is considered. The distributed auction algorithm (Algorithm~\ref{algo:disAlgo}) obtains very similar results as the centralized one (and as we have proved would converge to the same result, if the number of auction iterations would be large enough). We note here that 1000 iterations are considered to keep the algorithm real-time on our machine. Interestingly, the distributed auction yields a more egalitarian solution in terms of serviced customers by the three companies. 

\item Low communication heuristics (rows 3-4): the distributed auction algorithm is run with $500$ and $250$ iterations. This is to evaluate the degradation of performance with the decrease of the number of iterations: in the distributed auction, fewer iterations mean that fewer customers get assigned. For the simulation with $500$ iterations, the service rate decreases by $6.5$\%, and waiting and detour times are not afftected. When iterations are limited to $250$, detour times are lowered by $7.7\%$, at the price of a drop in service rate by $8.3\%$.

\item Stochasticity (rows 5-6): the travel times $c_{ij}$ are added a zero-mean Gaussian noise scalar with standard deviation indicated in the table (the noise is independent and identically distributed (IID) over the $c_{ij}$). This is to model errors in the cost computations, and therefore to evaluate the robustness of the algorithm to over/under-estimations of the travel times. With respect to the baseline, the waiting times increase by $28.13\%$ and	$34.38\%$, the detour times increase by $19.23\%$ and	$34.62\%$, however the SR degrades only slightly ($0.4$--$1$\%). In all these simulations, the smallest company is able to serve more customers than in the baseline, both in a relative and absolute way, and this, in turn, degrades the overall performance.  

\item Bias (rows 7-8): assignment costs are discounted by a percentage indicated in the table, for the smaller company (under-estimator bias). In particular, a $20\%$ reduction means that the respective cost in the linear assignment problem is $0.8$ times the original cost: the bias is considered in the assignment algorithm, but not in the actual implementation of the routing schedule, so the actual trip time is not affected. A bias introduced in this way simulates situations such as: a company entering in the market by discounting the rides to attract more clients, a city council that helps a new company to break in by waiving part of the new company rides, etc. The simulations show that the SR is seemingly not affected, the waiting and detour times decrease or do not change for the two bigger companies (having less customers to serve), but increase for the third company. The customer share increases slightly for the third company (both in relative and absolute terms), and the customer assignment share is similar to the centralized solution. This increase in shares enables to handle a bit more than 1000 additional rides per day
(in the 20\% discount simulation): this suggests that the bias variant may be a worthy choice, especially because the performance in terms of waiting and detour times degrades slightly.   
\end{enumerate}

In brief, the \Coop~model appears promising for reaching an egalitarian partitioning of the customers and, if a sufficient number of iterations are run, it obtains a performance similar to the centralized model. It is fairly robust to noise (both when travel times are over and under estimated), and it accommodates market incentives, like discounting prices to attract more customers for small companies.

\paragraph{\texttt{Competitive} model.} The performance assessment of the \Comp~model is presented in \Cref{tab:benchmark-comp}. The maximum number of iterations $k_{\text{comp}}$ is set to $17$.

\setlength\tabcolsep{4.0pt}
\begin{table}
\scriptsize
\centering
\caption{Simulations for the \Comp~scenario with three companies, having $265, 175, 60$ vehicles, respectively which correspond to a 53\%, 35\%, 12\% share. The two rows for the \textit{customers} column represent the difference in percentage w.r.t. an egalitarian partition of the serviced customers and the total percentage of the serviced customers, respectively. Waiting and detour times are reported both in cumulative and per-company values.}
\label{tab:benchmark-comp}
\begin{tabular}{ccccccccccc}
\toprule
sim. &  pref. & stoch & bias  & SR & customers & waiting &  detour    & occupancy & comp. time\\
& [-] & [\%] & [min] & [\%] & [\%] & [\%]   &  [min]   & [cust./vehicle] & [s] \\
\toprule
\rowcolor{Gray} 
1 &  Centr. & 0,0,0 & 0,0,0 & 99.4 & \mul{-2.3,0.8,1.5}{50.4, 35.6, 13.4} & \mul{3.6}{3.6, 3.6, 3.6} & \mul{2.9}{2.8, 2.9, 3.2}  & 0.7, 0.8, 0.9 & 1.5\\
\midrule
2 & 0 & 0,0,0 & 0,0,0 & 99.5 & \mul{-4.9,-2.3,7.2}{47.8, 32.5, 19.2} & \mul{3.7}{3.5, 3.5, 4.5} & \mul{3.1}{2.7, 2.8, 3.9}  & 0.6, 0.7, 1.7 & 1.8\\
\midrule
\rowcolor{Gray} 
3 & 0 & 5,5,5 & 0,0,0 & 99.2 & \mul{-4.3,-0.6,4.9}{48.3, 34.1, 16.8} & \mul{4.1}{4.0, 4.1, 4.5} & \mul{3.4}{3.2, 3.3, 3.9}  & 0.7, 0.8, 1.5 & 1.8\\
4  & 0 & 10,10,10 & 0,0,0 & 98.9 & \mul{-3.9,0.2,3.7}{48.6, 34.8, 15.5} & \mul{4.3}{4.2, 4.3, 4.6} & \mul{3.7}{3.5, 3.6, 4.1}  & 0.7, 0.9, 1.5 & 1.7\\
\midrule
\rowcolor{Gray}
5  & 0.5$^a$ & 0,0,0 & 0,0,0 & 99.0 & \mul{-4.5,0.4,4.1}{48.0, 35.0, 16.0} & \mul{3.8}{3.6, 3.8, 4.4} & \mul{3.2}{3.0, 3.2, 3.7}  & 0.6, 0.8, 1.3 & 2.0 \\ 
6  & 0.5$^a$ & 0,0,0 & 0,0,20 & 99.0 & \mul{-4.2,-0.4,4.6}{48.3, 34.3, 16.4} & \mul{3.8}{3.5, 3.8, 4.5} & \mul{3.2}{2.8, 3.1, 4.0}  & 0.6, 0.7, 1.5 & 1.8\\
\rowcolor{Gray}
7  & 1.$^a$ & 0,0,0 & 0,0,0 & 97.9 & \mul{-1.9,5.5,-3.6}{50.0, 39.6, 8.3}  & \mul{4.0}{3.7, 4.3, 3.8} & \mul{3.3}{3.1, 3.6, 3.0} & 0.7, 1.0, 0.5 & 1.8\\

8  & 1.$^a$ & 0,0,0 & 0,0,20 & 98.1 & \mul{-2.2,5.2,-3.0}{49.8, 39.4, 8.9} & \mul{3.9}{3.7, 4.3, 3.9} & \mul{3.3}{3.1, 3.6, 3.4}  & 0.7, 1.0, 0.6 & 1.8\\
\rowcolor{Gray}
9  & 0.5$^b$ & 0,0,0 & 0,0,0 & 98.9 & \mul{-4.7,0.3,4.4}{47.8, 34.9, 16.2} & \mul{3.8}{3.5, 3.8, 4.4} & \mul{3.2}{3.0, 3.3, 3.6}  & 0.6, 0.8, 1.3 & 2.1 \\
10  & 0.5$^b$ & 0,0,0 & 0,0,20 & 98.8 & \mul{-4.5,0.1,4.4}{47.9, 34.7, 16.2}  & \mul{3.7}{3.4, 3.7, 4.6} & \mul{3.3}{2.9, 3.2, 4.0} & 0.6, 0.8, 1.5 & 2.1\\
\rowcolor{Gray}
11  & 0.5$^b$ & 0,0,0 & 0,0,40 & 98.8  & \mul{-4.9,0.1,4.8}{47.5, 34.7, 16.6} & \mul{3.8}{3.5, 3.7, 4.6} & \mul{3.4}{2.9, 3.2, 4.3} & 0.6, 0.8, 1.7 & 2.0 \\
12  & 1.$^b$ & 0,0,0 & 0,0,0 & 96.8 & \mul{-1.3,5.9,-4.6}{50.1, 39.6, 7.1}  & \mul{3.9}{3.6, 4.4, 3.8} & \mul{3.4}{3.1, 3.7, 3.0} & 0.6, 1.1, 0.5 & 1.8 \\ 
\rowcolor{Gray}
13  & 1.$^b$ & 0,0,0 & 0,0,20 & 96.8 & \mul{-1.9,5.9,-4.0}{49.5, 39.6, 7.7} & \mul{4.0}{3.7, 4.4, 4.0} & \mul{3.4}{3.0, 3.7, 3.2}  & 0.7, 1.1, 0.5 & 1.7 \\ 
14  & 1.$^b$ & 0,0,0 & 0,0,40 & 97.1 & \mul{-2.8,6.0,-3.2}{48.7, 39.8, 8.6} & \mul{4.0}{3.7, 4.4, 4.1} & \mul{3.4}{3.0, 3.7, 3.5} & 0.6, 1.1, 0.6 & 1.8 \\
\midrule
\rowcolor{Gray}
15  & 0.95$^c$ & 0,0,0 & 0,0,0 & 92.9  & \mul{-3.1,6.8,-3.7}{46.3, 38.8, 7.8} & \mul{4.1}{3.9, 4.3, 3.7} & \mul{3.3}{3.0, 3.6, 3.0} & 0.6, 1.1, 0.5 & 1.6 \\ 
\toprule
\end{tabular}
\end{table}

\begin{enumerate}
\item Baseline (row 2): the \Comp~algorithm (Algorithm~\ref{algo:noCommAlgo}) without stochasticity, bias, or preferences. Note that each company solves the assignment problem with its fleet to optimality. Competition in its pristine form only slightly degrades the waiting and detour times significantly with respect to the centralized model, and it achieves a $0.1\%$ better SR. However, a suboptimal partition in terms of waiting and detour times is found, putting more load on the third company. This can be explained by the fact that the companies with a higher number of vehicles have higher chances to secure the best assignments (i.e., win their auctions), while the third company is left with the less attractive ones. Interestingly, the third company increases its market share, because of the context mapping. 
This simulation indicates that a certain level of competition does not deteriorate the service rate.

\item Stochasticity (rows 3-4): as in the \Coop~scenario, the travel times $c_{ij}$ are added a zero-mean Gaussian noise scalar with standard deviation indicated in the table (the noise is IID over the $c_{ij}$).  The SR degrades slightly (by $0.3$\% for simulation 3 and $0.6$\% for simulation 4), while the waiting times and detour times increase by $0.4$-$0.6$ and $0.3$-$0.6$ minutes, respectively. There is very little difference between the ``robustness'' of the \Comp~algorithm and the \Coop~one. Although the \Comp~algorithm generates lower variation in SR percentage, it is difficult to compare the variation in the share of customers assignment among companies.

\item Customer preferences (rows 5-15): since the third company seems to offer a worse service, customers may have a preference to be associated with the first two. We divide this study in three parts: low preference (indicated with $a$), medium preference ($b$), and high preference ($c$) settings, see Table~\ref{tab:benchmark-comp} (rows 5-15). Preferences are introduced in the simulation by giving the customers a certain propensity to choose a certain company despite the other companies offering a better ``deal'' (in terms of trip duration). A preference of $p$ in the table means that we set $100 p$\% of the customers to have a preference (preferences are divided equally between company 1 and company 2, since on average they offer a similar quality of service.

We also set the switching threshold, i.e., the maximum difference between the deal (i.e., trip duration) of their preferred company with the best deal of another company that would make them change their mind. This is to model the natural tendency of people to have a preference for a given provider, but be willing to switch to another one if what they get is better than a person-specific amount. We randomly assign the switching threshold to the customers by sampling from the vector $[s:5:30, 120]$ (vector expressed in MATLAB notation, and in minutes), so that some customer would be easy to switch, other less so. To say it in another way, e.g., if $s=5$, we decide the switching threshold of a customer by sampling one element from the vector $[5:5:30, 120] = [5,10,15,20,25,30,120]$ minutes, so that some customers may have low thresholds ($5$ minutes), others larger ones. 

In the simulations, $s$ is set to: 5 minutes for the low preference setting ($a$) (simulations 5-8), so more customers are willing to switch to another company; 15 minutes for the medium setting ($b$) (simulations 9-14); finally, 30 minutes for the high setting ($c$). This last setting is meant to model the realistic situation in which customers use their provider app and if a deal is not available that suits them, they would rather use a different means of transportation. We remark that the preferences override any other preference or hard constraints in terms of waiting and detour times.

The results show that introducing preferences has a detrimental effect on the overall performance, pushing down SR by $0.5$--$6.6$\% and pushing up the waiting and detour times. Also the equal division of preferences among the two bigger companies (i.e., the probability of preferring company 1 is equal to the probability of preferring company 2), puts a high workload on company 2 (having less vehicles than company 1), as shown by the increases of waiting and detour times. As natural for competitive settings, the third company has a more difficult time in securing customers and entering in the market. A bias term discounting the trips, as can be seen in simulations 6, 8, 10, 11, 13, 14, has a positive effect to counteract the preference competition, but this is limited by the fact that the third company is the one offering fewer vehicles.
\end{enumerate}

Another option that companies and cities have to augment their level of service is to increase the number of vehicles in service. 

This is explored in Table~\ref{tab:benchmarkci} (where simulations 1 and 2 are the centralized auction and the baseline \Comp~model, respectively). Here we increase the vehicles uniformly across companies, so an increase of $20\%$ means that the number of vehicles are $(265, 175, 60)\times 1.2$. We do not model the potential localized effects of congestion resulting from such an increase. In all the preference models, an increase of $10-20 \%$ of the fleet is required to obtain a similar performance (at least in terms of SR) to simulations 1 and 2. In the case of a high-preference model, i.e. customers booking with a company app, an increase of $40\%$ is instead needed. If such an increase is possible and not regulated, in a purely competitive market, this irremediably leads to more vehicles in cities, which in turn may jeopardize the early promises of ridesharing in terms of sizing down the number of vehicles. Note also that in a realistic scenario, not all the companies can leverage fleet sizing to increase their SR, which in turn may affect the market dynamics.

\setlength\tabcolsep{4.0pt}
\begin{table}
\scriptsize
\centering
\caption{Simulations for the \Comp~scenario with a varying number of vehicles, for each of the three companies. The baseline has $265, 175, 60$ vehicles, respectively which correspond to a 53\%, 35\%, 12\% share. The two rows for the \textit{customers} column represent the difference in percentage w.r.t. an egalitarian partition of the serviced customers and the total percentage of the serviced customers, respectively. Waiting and detour times are reported both in cumulative and per-company values. Stochasticity and bias are set to $0$.}
\label{tab:benchmarkci}
\begin{tabular}{cccccccccccc}
\toprule
sim. & vehicle increase & pref.  & SR & customers & waiting &  detour    & occupancy & comp. time\\
& [\%] & [-] &  [\%] & [\%]   &  [min]  & [\%] & [cust./vehicle] & [s] \\
\toprule
\rowcolor{Gray} 
1 & 0 & Centr. &  99.4 & \mul{-2.3,0.8,1.5}{50.4, 35.6, 13.4} & \mul{3.6}{3.6, 3.6, 3.6} & \mul{2.9}{2.8, 2.9, 3.2}  & 0.7, 0.8, 0.9 & 1.5\\
\midrule
2 & 0 & 0 & 99.5 & \mul{-4.9,-2.3,7.2}{47.8, 32.5, 19.2} & \mul{3.7}{3.5, 3.5, 4.5} & \mul{3.1}{2.7, 2.8, 3.9}  & 0.6, 0.7, 1.7 & 1.8\\
\midrule
\rowcolor{Gray}
3 & 0 & 1.$^a$ & 97.9 & \mul{-1.9,5.5,-3.6}{50.0, 39.6, 8.3} & \mul{4.0}{3.7, 4.3, 3.8} & \mul{3.3}{3.1, 3.6, 3.0}  & 0.7, 1.0, 0.5 & 1.8 \\
4 & +20\%  & 1.$^a$ & 99.7  & \mul{-3.8,9.7,-5.9}{49.0, 44.5, 6.2} & \mul{3.9}{3.8, 4.1, 3.7} & \mul{3.2}{3.0, 3.3, 3.1} & 0.6, 0.9, 0.3 & 2.4\\
\midrule
\rowcolor{Gray}
5 &0 & 1.$^b$  & 96.8 & \mul{-1.3,5.9,-4.6}{50.1, 39.6, 7.1}  & \mul{3.9}{3.6, 4.4, 3.8} & \mul{3.4}{3.1, 3.7, 3.0} & 0.6, 1.1, 0.5 & 1.8 \\ 
6 & +20\% & 1.$^b$  & 99.8 & \mul{-3.9,12.0,-8.1}{49.0, 46.9, 3.9} & \mul{3.8}{3.7, 3.9, 3.7} & \mul{3.2}{3.0, 3.3, 3.0}  & 0.5, 0.8, 0.2 & 3.0 \\
\midrule
\rowcolor{Gray}
7 &0 & 0.95$^c$  & 92.9 & \mul{-3.1,6.8,-3.7}{46.3, 38.8, 7.8} & \mul{4.1}{3.9, 4.3, 3.7} & \mul{3.3}{3.0, 3.6, 3.0}  & 0.6, 1.1, 0.5 & 1.6 \\ 
8 & +20\% & 0.95$^c$  & 98.4 & \mul{-5.1,9.8,-4.7}{47.1, 44.1, 7.2} & \mul{3.9}{3.8, 4.1, 3.5} & \mul{3.2}{3.0, 3.4, 2.9}  & 0.6, 0.9, 0.3 & 2.6 \\
\rowcolor{Gray}
9 & +40\% & 0.95$^c$ & 99.4 & \mul{-5.4,11.4,-6.0}{47.3, 46.1, 6.0} & \mul{3.7}{3.6, 3.8, 3.4} & \mul{3.1}{3.0, 3.2, 2.9}  & 0.5, 0.7, 0.2 & 3.2 \\
10 & +60\% & 0.95$^c$  & 100.0 & \mul{-4.7,10.9,-6.2}{48.3, 45.9, 5.8}  & \mul{3.6}{3.4, 3.8, 3.1} & \mul{3.0}{2.8, 3.2, 2.6} & 0.4, 0.6, 0.2 & 5.8 \\
\toprule
\end{tabular}
\end{table}

The increase in fleet size also has an adverse effect on the egalitarian partitioning, which is expected as the 2 preferred companies have more availability in terms of low occupancy vehicles that can pick up new customers. The \% of customers served by company 3 goes down to as low as  to 3.9\% in the case of 100\% of customers having a preference for company 1 or 2 with threshold b and 20\% more vehicles, and 5.8\% in the case 95\% of customers having a preference for company 1 or 2 with threshold c and 60\% more vehicles.

In brief, the competition of ridesharing companies may be as good as the centralized case in terms of the SR, however not as good in terms of average detour and waiting times, and in realistic settings where preferences are naturally present, it can lead to a significant loss in performance. In such a context, ridesharing companies may decide to significantly increase their operating fleet size to improve their service, to only get to a similar SR to the one achieved in the centralized and cooperative case without this fleet size increase.

\subsection{Main findings}\label{sec:num-findings}

The results highlighted in the previous Section~\ref{sec:num-real} should be put in perspective of the rise of Mobility as a Service platforms for city transport. We summarize them as follows:

\begin{itemize}

\item The \Coop~solution performs very closely to the \Centr~solution, is robust to noise, and responds well to incentives in market mechanisms. 

\item The \Coop~solution however requires communication with a MaaS platform and the agreements of companies to share bids at each iteration of the algorithm. Notice that our solution does not require the company to share proprietary information such as their vehicle locations but instead their best bids at each iteration, which could be implemented in a privacy-aware solution. 

\item The \Comp~solution achieves good performance in terms of SR but exhibits a deterioration in waiting and detour times compared to the \Centr~solution. 

\item The \Comp~solution also requires communication with a Maas platform but requires less iterations for convergence. It does require the sharing of assignments at each iteration of the algorithm. 

\item The \Comp~solution tends to favor the company with the lowest market share. 

\item Considering customers' preferences is a proxy for modeling either the possibility for customers to choose specific companies, or the absence of a MaaS platform, i.e. customers book through the proprietary smartphone applications of the preferred companies. The \Comp~model coping with customers' preferences sensibly deteriorates the SR, as well as the waiting and detour times.

\item The \Comp~solution, when considering customers' preferences, has a serious impact on the long term quality of ridesharing services, and an increase of the number of operating vehicles is needed to achieve similar service rate to the \Centr~solution.

\end{itemize}

In brief, we advocate for the needs to regulate the settings of MaaS platforms and the possibility for customers to choose their preferred companies. First, MaaS platforms must be specific about the information that is wanted from the ridesharing companies and a regulatory framework is needed to agree on the level of information sharing, keeping in mind the common good of ridesharing mobility and the privacy of the involved companies. Second, MaaS platforms cannot behave as plain brokers, leaving the final choice to the customers. The possibility for customers to have company preferences sensibly leads to sub-optimal performance of the system. Finally, a regulatory framework is needed for fleet sizing in a way that is fair to all companies involved, in order to avoid an unhealthy increase of supply by each company in order to gain market shares.

		

\section{Conclusions}	\label{sec:conclusion}

Shared-mobility services are facing an ever-growing trend in cities. The benefits of ridesharing may be jeopardized by an unregulated design and maintenance of those, especially in case multiple companies are competing for their market share.
In this work, we have presented optimization-based approaches to model cooperation and competition between ridesharing companies. A cooperative model, which could make use of Mobility as a Service platforms, is shown to solve to optimality the problem of assigning vehicles to customer request, by following closely results from the literature: in such a model, the distributed auction algorithm is exploited to limit the amount of information that the companies are required to share. We have investigated how a realistic model of competition deviates from this optimality and provided worst case bounds. We have evaluated the performance of both models in terms of service rate, waiting and detour times on one-week instances of the New York City taxi dataset. Model variants coping with noise in the travel time estimations, bias in the assignment costs, and preferences in the competitive case have been presented and validated. 
The computational results suggest that cooperation among ridesharing companies can be conducted in such a way to limit the degradation of the level of service with respect to a centralized model. The competition can instead lower the quality of the ridesharing service, especially in the case customer preferences are accommodated.
 Finally, we envision several research directions as future work: (i) formulating company-wise strategies to gain market share in a defavorable market (e.g., fine tuned strategies related to the spatio-temporal distribution of vehicles in response to the spatio-temporal SR per company achieved); (ii)  dynamic fleet sizing at the city level: what is the minimum of vehicles needed to achieve satisfactory SR, waiting and detour times.; (iii) model the congestion feedback effect caused by uncontrolled supply in competitive settings.

 \section{Acknowledgments}
This project has received funding from the European Union Horizon 2020 research and innovation programme under grant agreement No. 731993 (AUTOPILOT). 

\bibliography{Autopilot_Literature,NewReferences}

\end{document}